\documentclass[%
 reprint,
superscriptaddress,
 amsmath,amssymb,
 aps,
prc,
]{revtex4-2}

\usepackage{graphicx}% Include figure files
\usepackage{dcolumn}% Align table columns on decimal point

\usepackage{bm}
\usepackage{multirow}% bold math
\begin{document}

 \preprint{APS/123-QED}

\title{ CREX and PREX-II motivated  relativistic interactions and their implications to the bulk properties of nuclear matter and neutron star}
\author{Mukul Kumar}
\email{mukulpathania120495@gmail.com}
\affiliation{Department of Physics, Himachal Pradesh University, Shimla-171005, India}%
\author{Sunil Kumar}
 \affiliation{Department of Physics, Himachal Pradesh University, Shimla-171005, India}

  \author{Virender Thakur}
 \email{virenthakur2154@gmail.com}
 \affiliation{Department of Physics, Himachal Pradesh University, Shimla-171005, India}
\author{Raj Kumar}
\email{raj.phy@gmail.com}
\affiliation{Department of Physics, Himachal Pradesh University, Shimla-171005, India}

\author{B.K. Agrawal}
\email{sinp.bijay@gmail.com}
\affiliation{Saha Institute of Nuclear Physics, 1/AF Bidhannagar, Kolkata 700064, India}

\author{Shashi K. Dhiman}
\email{shashi.dhiman@gmail.com}
\affiliation{Department of Physics, Himachal Pradesh University, Shimla-171005, India}
\affiliation{School of Applied Sciences, Himachal Pradesh Technical University, Hamirpur-177001, India}

\begin{abstract}
 We investigate the implications of parity-violating electron scattering experiment on  neutron skin thickness  of  $^{48}$Ca (CREX) and $^{208}$Pb (PREX-II)  data on the bulk  properties of  finite nuclei, nuclear matter and neutron stars. The neutron skin thickness from the CREX and PREX-II data is employed to constrain the parameters  of relativistic mean field models which includes different non-linear, self and cross-couplings among isoscalar-scalar $\sigma$, isoscalar-vector $\omega$, isovector-scalar $\delta$ and isovector-vector $\rho$ meson fields up to the quartic order. Three  parametrizations of RMF model are proposed by fitting CREX, PREX-II and both
CREX as well as PREX-II data to  assess their implications. A covariance analysis is performed to assess the theoretical uncertainties  of model parameters and nuclear matter observables along with correlations among them. The RMF model parametrization obtained  with the CREX data acquires much smaller value of symmetry energy (J= 28.97$\pm$ 0.99 MeV), its slope parameter  (L= 30.61$\pm 6.74$ MeV)  in comparison to those obtained with  PREX-II data. The  neutron star properties are studied by employing the equations of state (EoSs) composed of nucleons and leptons in $\beta$ equilibrium.
\end{abstract}
%\PACS{97.60.Jd; 26.60.+c; 21.65.+f}
\maketitle

%\tableofcontents

\section{INTRODUCTION}
The nuclear equation of state (EoS) plays a vital role  for understanding the properties of strongly interacting many body systems like atomic nuclei and neutron stars \cite{Lattimer2012,Oertel2017,Rocamaza2018}. The nuclear symmetry energy  and its density dependence are key features of nuclear EoS. Constraining the density dependence  of symmetry energy represents a long-standing and unresolved question in nuclear physics and astrophysics \cite{Rocamaza2018}. The  density dependence of the symmetry energy has implications in a variety of phenomena such as heavy-ion collisions, core-collapse supernovas and neutron-star structures. Although important, this quantity cannot be directly measured in the laboratory, it can only be derived from theories and thus to constrain their values it is necessary to identify and use relevant observables on finite nuclei.
A neutron star is a highly dense and asymmetric nuclear system that has  a central density of about 5-6 times the nuclear saturation density \cite{Lattimer2004}. The study of the neutron star proclaims that its internal structure is more complex as new degrees of freedom like hyperons and quarks appear in the core. The properties of the neutron star like mass, radius, and tidal deformability can be explained by taking into account the interaction between nucleons and the mesonic degree of freedom in the form of Lagrangian. This provides an EoS  which is the main input for the calculation of neutron star properties. The several relativistic mean fields (RMF) models  having effective lagrangian density consisting of nonlinear $\sigma$, $\omega$, $\rho$ and $\delta$  terms and cross terms have been analyzed for nucleonic matter  and nucleonic along with hyperonic matter and accosted with the constraints of nuclear matter properties and astrophysical observations of compact star masses \cite{Dhiman2007,Virender2022a,Virender2022b}.
 The nuclear theory studies \cite{Haensel2007,Lattimer2014,Baym2018} 
are mainly focusing on understanding the dense matter in compact stars (CS). The constraints on EoS at high density are imposed with accurate information of a neutron star's maximum mass and radius \cite{Hebeler2010,Hebeler2013,Lattimer2012}. The precise measurement of masses of millisecond pulsars such as PSR J1614-2230 \cite{Demorest2010}, PSR J0348+0432 \cite{Antoniadis2013} show that the maximum mass of the neutron star should be around 2 M$\odot$.
The recent observations with LIGO and Virgo of GW170817 event \cite{Abbott2018,Abbott2019} of Binary Neutron Stars merger
and the discovery of CS with masses around 2$M_\odot$ \cite{Demorest2010,Antoniadis2013,Arzoumanian2018,Miller2019,Riley2019,Raaijmakers2019} have intensified the interest in these 
intriguing objects. The analysis of GW170817 has demonstrated the potential
of gravitational wave (GW) observations to yield new information relating to 
the limits on CS tidal deformability.\\
The recent precise parity-violating electron scattering experiments on $^{48}$Ca (CREX) \cite{Adhikari2022} and $^{208}$Pb (PREX-II)\cite{Adhikari2021} provide new insights into the neutron skin thickness of nuclei. These experiments are helpful in determining the nuclear weak charge form factor by measuring the parity-violating asymmetry. The weak charge form factor has a strong correlation with the density dependence of symmetry energy and neutron skin thickness of nuclei and plays an important role in probing the isovector channels of energy density functionals. The parity-violating electron scattering experiments give precise and model-independent data for the nuclear weak  charge form factor that can be used to constrain the energy density functionals \cite{Reinhard2013}. The weak charge form factors   of $^{48}$Ca and $^{208}$Pb reported by  CREX and PREX-II experiments and  their  measured   parity-violating asymmetry  has been analyzed using density functionals and reached on a conclusion that it is difficult to describe parity-violating asymmetry  simultaneously in both nuclei    \cite{Reinhard2021,Reinhard2022}. The Calcium Radius Experiment (CREX) has recently given a model-independent extraction of neutron skin thickness of $^{48}{Ca}$ as $\Delta r_{np}$ = 0.121 $\pm$ 0.026 fm \cite{Adhikari2022} which suggests  softness of density dependence of symmetry energy.
The Lead Radius Experiment (PREX) has recently given a model-independent extraction of neutron skin thickness of $^{208}{Pb}$ as $\Delta r_{np}$ = 0.283 $\pm$ 0.071 fm \cite{Adhikari2021} by combining the original PREX result with the new PREX-II. The $\Delta r_{np}$ has been identified as an ideal probe on symmetry energy - a key but poorly known quantity that describes the isospin dependence of EoS of nuclear matter and plays a crucial role in various issues in nuclear physics and astrophysics. The neutron skin thickness of the Lead nucleus exhibits a strong positive linear correlation with the slope of  symmetry energy (L) at saturation density. The value of  L around the saturation density strongly affects the Mass-Radius relation and tidal deformability ($\Lambda$) of a neutron star and provides a unique bridge between atomic nuclei and neutron star. The large value of  $\Delta r_{np}$ = 0.283 $\pm$ 0.071 fm suggests a very stiff EoS and large value of L around saturation density and generally gives rise to a large value of neutron star radius and the tidal deformability \cite{Reed2021}. The upper limit on $\Lambda_{1.4}$ $\leq$ 580 for GW170817 requires softer EoS and hence softer symmetry energy coefficient \cite{Abbott2018}. The heaviest neutron star 2.14$_{-0.09}^{+0.10}M_{\odot}$ of PSRJ0740+6620 \cite{Cromartie2020} also limits the EoS for symmetric nuclear matter (SNM). The flow data from heavy ion collisions  suggests that the EoS for SNM should be  relatively softer \cite{Danielewicz2002}.\\
The motivation of the present work is to generate three  new relativistic interactions for the Lagrangian density of the  RMF model  to investigate the effect of CREX and PREX-II data on neutron skin thickness for $^{48}$Ca and  $^{208}$Pb nuclei on the bulk nuclear properties and  observed astrophysical constraints on neutron stars.  A covariance analysis can also be performed to assess the theoretical uncertainties  of model parameters and nuclear matter observables along with correlations among them. \\
The paper is organized as follows, in section \ref{tm}, a brief outlines of the RMF Lagrangian, equations of motion and  EoS for neutron stars is provided.
In section \ref{oaca}, the procedure for optimization  of the model parameters and covariance analysis is discussed. Numerical results and detailed discussions features of model  parametrizations,  finite nuclei, bulk nuclear matter, neutron star matter and correlations amongst nuclear matter observables and model parameters are presented in section \ref{results}. Finally, we give a summary in section \ref{summary}.
\section{THEORETICAL MODEL}\label{tm}
The Lagrangian  density  for  the RMF model used in the present work is  based upon
different non-linear, self and inter-couplings among isoscalar-scalar $\sigma$,
isoscalar-vector $\omega_{\mu}$, isovector-scalar $\delta$ and isovector-vector
$\rho_{\mu}$ meson fields and nucleonic Dirac field $\Psi$ \cite{Dhiman2007, Raj2006,Virender2022a,Virender2022b},
is given by
\begin{eqnarray}
\label{eq:lbm}
{\cal L} &=& \sum_{B} \overline{\Psi}_{B}[i\gamma^{\mu}\partial_{\mu}-
(M_{B}-g_{\sigma B} \sigma- g_{\delta B}\delta\cdot\tau_{3})-(g_{\omega B}\gamma^{\mu} \omega_{\mu}\nonumber\\&+&
\frac{1}{2}g_{\mathbf{\rho}B}\gamma^{\mu}\tau_{B}.\mathbf{\rho}_{\mu})]\Psi_{B}
+ \frac{1}{2}(\partial_{\mu}\sigma\partial^{\mu}\sigma-m_{\sigma}^2\sigma^2)\nonumber\\  &-&
\frac{\overline{\kappa}}{3!}
g_{\sigma N}^3\sigma^3-\frac{\overline{\lambda}}{4!}g_{\sigma N}^4\sigma^4  - \frac{1}{4}\omega_{\mu\nu}\omega^{\mu\nu}
+ \frac{1}{2}m_{\omega}^2\omega_{\mu}\omega^{\mu}\nonumber\\&+& \frac{1}{4!}\zeta g_{\omega N}^{4}(\omega_{\mu}\omega^{\mu})^{2}-\frac{1}{4}\mathbf{\rho}_{\mu\nu}\mathbf{\rho}^{\mu\nu}+\frac{1}{2}m_{\rho}^2\mathbf{\rho}_{\mu}\mathbf{\rho}^{\mu}\nonumber\\
&+&\frac{1}{2}(\partial_{\mu}\delta\partial^{\mu}\delta-m_{\delta}^2\delta^2)
	+\frac{1}{2}\Lambda_{\omega\rho}g_{\omega }^{2}g_{\rho }^2\omega_{\mu}\omega^{\mu}\rho_{\mu}\rho^{\mu}\nonumber\\
&-& \frac{1}{4}F_{\mu\nu}F^{\mu\nu}- \sum_{B}e\overline{\Psi} _{B}\gamma_{\mu}\frac{1+\tau_{3B}}{2}A_{\mu}\Psi_{B}\nonumber\\
&+&\sum_{\ell=e,\mu}{\overline{\Psi}_{\ell}}\left(i\gamma^{\mu}\partial_{\mu} -
M_{\ell}\right) \Psi_{\ell}.
\end{eqnarray}
% \begin{equation}
% \label{eq:lem}
% {\cal L}_{em}= -\frac{1}{4}F_{\mu\nu}F^{\mu\nu}- \sum_{B}e\overline{\Psi} _{B}\gamma_{\mu}\frac{1+\tau_{3B}}{2}A_{\mu}\Psi_{B},
% \end{equation}
% \begin{equation}
% \label{eq:lemu}
% {\cal L}_{e\mu}=\sum_{\ell=e,\mu}{\overline{\Psi}_{\ell}}\left(i\gamma^{\mu}\partial_{\mu} -
% M_{\ell}\right) \Psi_{\ell}.
% \end{equation}
The equation of motion for baryons, mesons, and photons can be derived from the Lagrangian
density defined in Eq.(\ref{eq:lbm}). The equation of motion for baryons can be given as,

\begin{eqnarray}
\label{eq:dirac}
&\bigg[\gamma^\mu\left(i\partial_\mu - g_{\omega B}\omega_\mu-\frac{1}{2}g_{\rho
B}\tau_{B}.\rho_\mu - e \frac{1+\tau_{3B}}{2}A_\mu \right) - \nonumber\\
&  (M_B - g_{\sigma B}\sigma - g_{\delta B}\delta\cdot\tau_{3})\bigg]\Psi_B
=\epsilon_B \Psi_B.
\end{eqnarray}

The Euler-Lagrange equations for the ground-state expectation values of the mesons fields are 

\begin{eqnarray}
\label{eq:sigma}
\left(-\Delta + m_{\sigma}^{2}\right)\sigma & = &\sum_{B} g_{\sigma B}\rho_{sB} -\frac{ \overline{\kappa}}{2} g_{\sigma N}^{3}\sigma^{2}- \frac{\overline{\lambda}}{6} g_{\sigma N}^{4}\sigma ^{3} \nonumber \\&&
%+\frac{\partial}{\partial \sigma} {\cal L}_{\sigma\omega\rho}
%{a_{1}} g_{\sigma N} g_{\omega N}^{2}\omega ^{2}
 %+a_{2} g_{\sigma N}^{2}
 %g_{\omega N}^{2}\sigma\omega ^{2}
 %\nonumber\\&&+ b_{1} g_{\sigma N} g_{\rho B}^{2}\rho ^{2}
 %+ b_{2} g_{\sigma N}^{2}
%g_{\rho N}^{2}\sigma\rho ^{2},
\end{eqnarray}
\begin{eqnarray}
\label{eq:omega}
\left(-\Delta + m_{\omega}^{2}\right)\omega & = &\sum_{B} g_{\omega B}\rho_{B} 
- \frac{\zeta}{6} g_{\omega N}^{4}\omega ^{3}\nonumber\\
%-\frac{\partial}{\partial \omega} {\cal L}_{\sigma\omega\rho} 
%\nonumber\\&&-2 a_{1}g_{\sigma N} g_{\omega N}^{2}\sigma\omega
%-a_{2} g_{\sigma N}^{2}
% g_{\omega N}^{2}\sigma^{2}\omega \nonumber\\
&-& \Lambda_{\omega\rho} g_{\omega N}^{2}
g_{\rho N}^{2}\omega\rho ^{2}, 
\end{eqnarray}
\begin{eqnarray}
\label{eq:rho}
\left(-\Delta + m_{\rho}^{2}\right)\rho & = &\sum_{B} g_{\rho B}\tau_{3B}\rho_{B}- \frac{\xi}{6} g_{\rho N}^{4}\rho ^{3} 
%-\frac{\partial}{\partial \sigma} {\cal L}_{\sigma\omega\rho}
\nonumber\\ &&
- \Lambda_{\omega\rho} g_{\omega N}^{2}
g_{\rho N}^{2}\omega^{2}\rho,  
\end{eqnarray}
\begin{equation}
\label{eq:delta}
\left(-\Delta + m_{\delta}^{2}\right)\delta  = \sum_{B} g_{\delta B}\rho_{s3B}
\end{equation}
\begin{equation}
\label{eq:photon}
-\Delta A_{0} = e\rho_{p}.
\end{equation}
where the baryon vector density $\rho_B$,  scalar density $\rho_{sB}$ and charge density
$\rho_{p}$ are, respectively, 
\begin{equation}
\rho_{B}= \left< \overline{\Psi}_B \gamma^0 \Psi_B\right> = \frac{\gamma k_{B}^{3}}{6\pi^{2}},
\end{equation}

\begin{equation}
\rho_{sB} = \left< \overline{\Psi}_B\Psi_B \right> 
          = \frac{\gamma}{(2\pi)^3}\int_{0}^{k_{B}}d^{3}k \frac{M_{B}^*}
            {\sqrt{k^2 + M_{B}^{*2}}},
\end{equation}
\begin{equation}
\rho_{p} = \left< \overline{\Psi}_B\gamma^{0}\frac{1+\tau_{3B}}{2}\Psi_B \right>, 
\end{equation}
 with $\gamma$  the spin-isospin degeneracy.
 The Dirac effective mass for the neutron and proton can be written as
\begin{equation}
\label{eq:nucleon}
	M^{*}_{p} =( M - g_{\sigma}\sigma - g_{\delta}\delta),
\end{equation}
\begin{equation}
\label{eq:nucleon1}
  M^{*}_{n} =( M - g_{\sigma}\sigma + g_{\delta}\delta),
\end{equation}
Following the Euler-Lagrange
formalism one can readily find the expressions for energy density ${\cal E}$ and
pressure $P$ as a function of density from Eq. (\ref{eq:lbm}) \cite{Glendenning2000}.

The energy density of the uniform matter  within the framework of the RMF model is given by;
\begin{equation}
\label{eq:eden}
\begin{split}
{\cal E} & = \sum_{j=B,\ell}\frac{1}{\pi^{2}}\int_{0}^{k_j}k^2\sqrt{k^2+M_{j}^{*2}} dk\\
&+\sum_{B}g_{\omega B}\omega\rho_{B}
+\sum_{B}g_{\rho B}\tau_{3B}\rho_{B}\rho
+ \frac{1}{2}m_{\sigma}^2\sigma^2\\
&+\frac{\overline{\kappa}}{6}g_{\sigma N}^3\sigma^3
+\frac{\overline{\lambda}}{24}g_{\sigma N}^4\sigma^4
-\frac{\zeta}{24}g_{\omega N}^4\omega^4\\
&-\frac{\xi}{24}g_{\rho N}^4\rho^4
 - \frac{1}{2} m_{\omega}^2 \omega ^2
-\frac{1}{2} m_{\rho}^2 \rho ^2\\
 &- \frac{1}{2} \Lambda_{\omega\rho} g_{\omega N}^2 g_{\rho N}^2
\omega^2\rho^2  + \frac{1}{2}m_{\delta}^2\delta^2\\
\end{split}
\end{equation}
The pressure of the uniform matter  is given by
\begin{equation}
\label{eq:pden}
\begin{split}
P & = \sum_{j=B,\ell}\frac{1}{3\pi^{2}}\int_{0}^{k_j}
\frac{k^{4}dk}{\sqrt{k^2+M_{j}^{*2}}} 
- \frac{1}{2}m_{\sigma}^2\sigma^2\\
&-\frac{\overline{\kappa}}{6}g_{\sigma N}^3\sigma^3 
-\frac{\overline{\lambda}}{24}g_{\sigma N}^4\sigma^4
+\frac{\zeta}{24}g_{\omega N}^4\omega^4\\
&+\frac{\xi}{24}g_{\rho N}^4\rho^4
  + \frac{1}{2} m_{\omega}^2 \omega ^2
+\frac{1}{2} m_{\rho}^2 \rho ^2 \\
&+ \frac{1}{2} \Lambda_{\omega\rho} g_{\omega N}^2 g_{\rho N}^2
\omega^2\rho^2  - \frac{1}{2}m_{\delta}^2\delta^2\\
\end{split}
\end{equation}
Here, the sum is taken over nucleons and leptons.
\section{OPTIMIZATION AND COVARIANCE ANALYSIS} \label{oaca}
The optimization of the parameters ($\textbf{p}$) appearing in the Lagrangian (Eq. \ref{eq:lbm}) has been performed  by using the simulated annealing method (SAM) \cite{Burvenich2004, 
Kirkpatrick1984} by following $\chi^{2}$ minimization procedure  which is given
as, 
\begin{equation} 
	{\chi^2}(\textbf{p}) =  \frac{1}{N_d - N_p}\sum_{i=1}^{N_d}
\left (\frac{ M_i^{exp} - M_i^{th}}{\sigma_i}\right )^2 \label {chi2},
\end{equation}
where $N_d$ is the number of  experimental data points and $N_p$ is the number 
of fitted parameters. The $\sigma_i$ denotes adopted errors \cite{Dobaczewski2014}
and $M_i^{exp}$ and $M_i^{th}$ are the experimental and the corresponding
theoretical values, respectively, for a given observable.
The minimum value of
${{\chi}}^{2}_{0}$  corresponds to the optimal values $\bf{p}_{0}$ of the parameters.
After  the optimization of the energy density functional, it is important to explore the richness of the covariance analysis. It enables one to calculate the statistical uncertainties/errors on model parameters or any calculated physical observables. The covariance analysis  also provides additional information about the sensitivity of the parameters to the physical observables, and interdependence among the parameters \cite{Dobaczewski2014,Chen2015,Mondal2015,Fattoyev2011}.
Having obtained the  parameter set, the correlation coefficient
between two quantities Y and Z can be calculated by covariance analysis
\cite{Brandt1997,Reinhard2010,Fattoyev2011,Dobaczewski2014,Mondal2015} as
\begin{equation}
\label{covariance}
    \textit{c}_{YZ} = \frac{\overline{\Delta{Y} \Delta{Z} }}{\sqrt{\overline{\Delta{Y^2}}
  \quad \overline{\Delta{Z^2}}}} ,
\end{equation}
where covariance between Y and Z is expressed as
\begin{equation}
\label{error}
    \overline{\Delta{Y}\Delta{Z}} = \sum_{\alpha\beta} \left( \frac{\partial{Y}}
{\partial{p}_{\alpha}}\right) _{\textbf{p}_0} C_{\alpha\beta}^{-1}
\left( \frac{\partial{Z}}{\partial{p}_{\beta}}\right) _{\textbf{p}_0}.
\end{equation}
Here, $C_{\alpha\beta}^{-1}$ is an element of inverted curvature matrix given by
\begin{equation}
\label{matrix}
    \textit{C}_{\alpha\beta} = \frac{1}{2}\left(\frac{\partial^2
	\chi^2(\textbf{p})}{\partial{p}_{\alpha}\partial{p}_{\beta}}\right)_{\textbf{p}_{0}}.
\end{equation}
The standard deviation, $\overline{\Delta{Y}^2}$, in Y can be computed using
Eq. (\ref{error}) by substituting Z = Y.
\section{RESULTS and DISCUSSION}\label{results}
We obtain parameterizations  for RMF models by employing CREX, PREX-II and combined CREX -PREX II data by following the procedure discussed in Section \ref{oaca}. The  model parameterizations  obtained are then used  to calculate the properties of finite nuclei and infinite nuclear matter and neutron stars. We also discuss the correlations among nuclear matter observables and model parameters.
\subsection{Parametrizations of RMF Model} \label{par}
\begin{figure}
\includegraphics[trim=0 0 0 0,clip,scale=0.51]{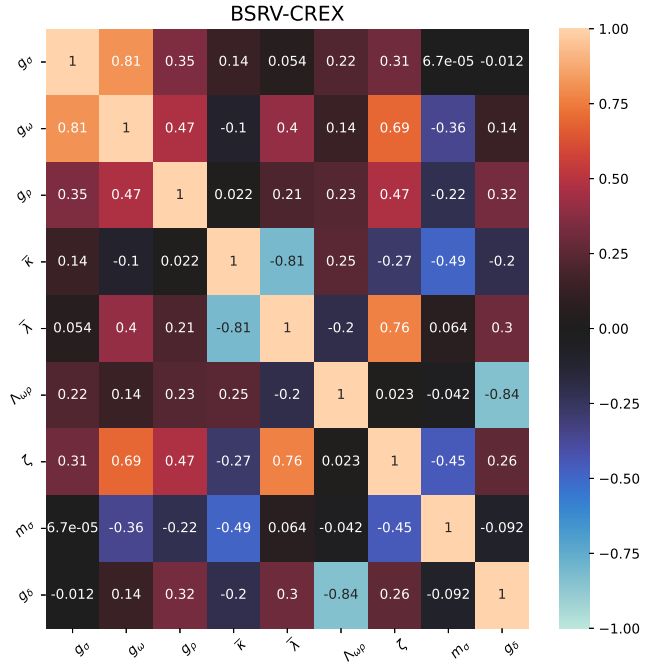}
\caption{\label{par_crex} (color online)  Correlation coefficients   among the model parameters
of the Lagrangian given by Eq. (\ref{eq:lbm}) for BSRV-CREX parametrization.  }
\end{figure}

\begin{figure}
\includegraphics[trim=0 0 0 0,clip,scale=0.51]{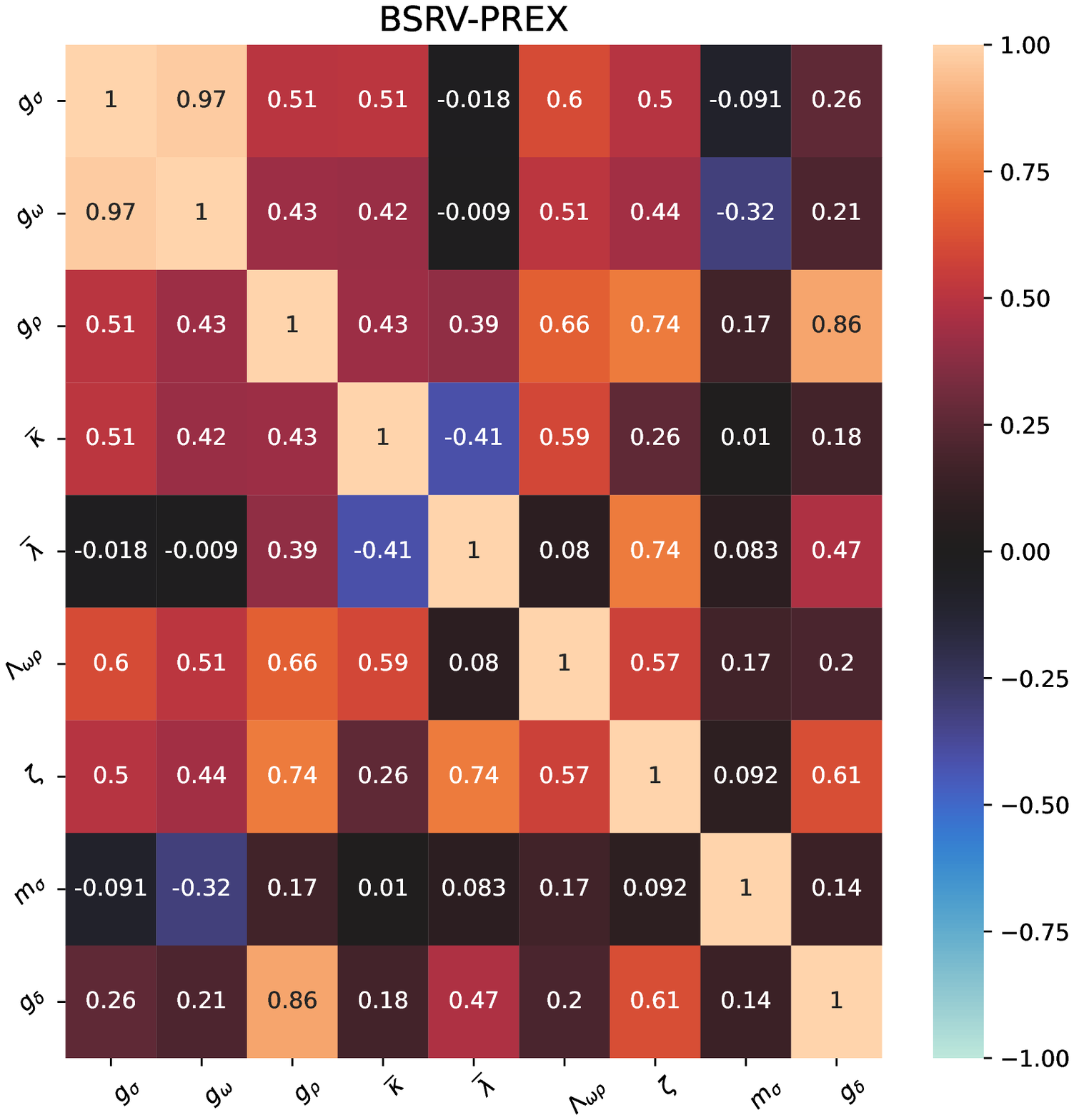}
\caption{\label{par_prex} (color online) Same as Fig. \ref{par_crex}, but for BSRV-PREX parametrization.}
\end{figure}

\begin{figure}
\includegraphics[trim=0 0 0 0,clip,scale=0.5]{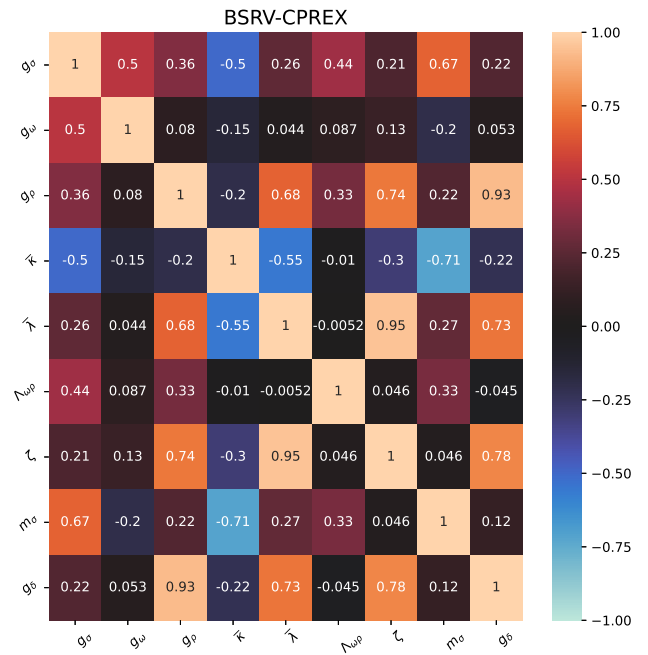}
\caption{\label{par_cprex} (color online) Same as Fig. \ref{par_crex}, but for for BSRV-CPREX parametrization). }
\end{figure}

In the present study, three  new relativistic interactions BSRV-CREX, BSRV-PREX and BSRV-CPREX  have been generated for the Lagrangian density
given by Eq. (\ref{eq:lbm}) to investigate the effect of CREX and PREX-II data on neutron skin thickness for $^{48}Ca $ and  $^{208}Pb$
nuclei on the properties of finite nuclei and neutron star matter.  The parameters of the BSRV-CREX, BSRV-PREX and BSRV-CPREX models are obtained by fitting exactly the available experimental data of
\cite{Wang2021} on binding energies ($BE$) and charge rms radii ($r_{ch}$) \cite{Angeli2013} of some closed/open-shell nuclei $^{16,24}$O, $^{40,48}$Ca, $^{56,68,78}$Ni,
$^{88}$Sr,$^{90}$Zr, $^{100,116,132}$Sn, $^{144}$Sm and $^{208}$Pb. In addition, we also include in our fit the value of neutron skin thickness for $^{48}$Ca and $^{208}$ Pb  nuclei, which is a very important physical observable to constrain the value of L that determines the linear dependence of symmetry energy. \\
\begin{table*}
\centering
%\begin{table}[p]
\caption{\label{tab:table1}
Newly generated parameter sets BSRV-CREX, BSRV-PREX and BSRV-CPREX  for  the Lagrangian of RMF model as given
in Eq.(\ref{eq:lbm}) along with theoretical uncertainties/errors. The parameters $\overline{\kappa}$, is
 in  fm$^{-1}$. The mass for nucleon,  $\omega$, $\rho$ and $\delta$ mesons  are taken as $M_N$ = 939 MeV, $m_{\omega}$= 782.5 MeV, $m_{\rho}$= 762.468 MeV and $m_{\delta}$= 980 MeV respectively. The values of $\overline{\kappa}$, $\overline{\lambda}$,
and ${\Lambda_{\omega\rho}}$ are multiplied by $10^{2}$. Parameters for NL3 \cite{Lalazissis1997}, FSUGarnet \cite{Chen2015}, IOPB-1 \cite{Kumar2018}, and Big Apple \cite{Fattoyev2020} are also shown for comparison.}
%\begin{ruledtabular}
\vskip 1cm
\begin{tabular}{ccccccccc}
\hline
\hline
\multicolumn{1}{c}{${\bf {Parameters}}$}&
\multicolumn{1}{c}{{\bf BSRV-CREX}}&
\multicolumn{1}{c}{{\bf BSRV-PREX}}&
\multicolumn{1}{c}{{\bf BSRV-CPREX}}&
\multicolumn{1}{c}{{\bf NL3}}&
\multicolumn{1}{c}{{\bf FSUGarnet}}&
\multicolumn{1}{c}{{\bf IOPB-1}}&
\multicolumn{1}{c}{{\bf Big Apple}}\\
\hline
${\bf g_{\sigma}}$& 10.71506$\pm$ 0.02086 &10.40613 $\pm$0.08977&10.44537$\pm$0.02480&10.21743 &10.50315&10.41851&9.67810\\
${\bf g_{\omega}}$  &13.82692$\pm$0.03539&13.37605$\pm$0.13078&13.43408$\pm$0.03477 & 12.86762&13.69695&13.38412&12.33541\\
${\bf g_{\rho}}$ &16.18406$\pm$1.20866&10.27951$\pm$2.23787&10.28003$\pm$2.15232&8.94880 &13.87880 &11.11560&14.14256\\
${\bf g_{\delta}}$ &4.27816$\pm$0.75631&1.19517$\pm$4.29242&1.70339$\pm$3.97826&--&-- &--&--\\
${\bf \overline {\kappa}}$&1.41726$\pm$0.04886 &1.64259$\pm$0.04079&1.66238$\pm$0.05821&1.95734 &1.65229 &1.85581&2.61776\\
${\bf \overline {\lambda}}$&0.46733$\pm$0.08773&-0.08316$\pm$0.09341&-0.20868$\pm$0.19224&-1.59137&-0.035330& -0.075516&-2.16586\\
${\bf{\zeta}}$&0.03441$\pm$0.00130&0.02611$\pm$0.00212&0.02429$\pm$0.00326&0.00000&0.23486& 0.017442&0.000699\\
${\bf \Lambda_{\omega\rho}}$ &5.18286$\pm$1.21735&2.90293$\pm$2.60556&2.25722$\pm$1.70126&0.00000 &8.6000 &4.80000&9.40000\\
${\bf m_{\sigma}}$ &504.679$\pm$0.689&502.050$\pm$1.099&501.933$\pm$1.600&508.194 &496.731 &500.487&492.975\\
\hline
\hline
\end{tabular}
%\end{ruledtabular}
%\end{table}
\end{table*}
The BSRV-CREX parametrization has been obtained by incorporating the recently measured neutron skin thickness $\Delta r_{np}$= $0.121\pm 0.026$ fm for $^{48}$Ca using the parity-violating electron scattering experiment \cite{Adhikari2022}. The parameters of BSRV-PREX model have been searched by incorporating the recently measured neutron skin thickness $\Delta r_{np}$= $0.283\pm 0.071$ fm for $^{208}$Pb from the PREX-II data  \cite{Adhikari2022} in our fit. The BSRV-CPREX parametrizations has been obtained by incorporating both the CREX and PREX-II data for neutron skin thicknesses for $^{48}$Ca and $^{208}$Pb nuclei in the fit data. In addition, we have also included the maximum mass of neutron star \cite{Rezzolla2018} in our fit data.\\
For the open shell nuclei, the pairing has been included using BCS formalism  with constant pairing gaps \cite{Ring1980,Karatzikos2010} that are taken from the nucleon separation energies  of neighboring nuclei \cite{Wang2021}. Neutron and proton pairing gaps  are calculated by using  the fourth-order finite difference mass formula (five-point difference) \cite{Duguet2001}. The neutron and proton pairing gaps ($\Delta_{n}$,$\Delta_{p}$)  in  MeV for the open shell nuclei  are $^{68}Ni$(1.46,0.0),  $^{88}Sr$(0.0,1.284), $^{90}Zr$(0.0,1.239), $^{116}Sn$(1.189,0.0) and $^{144}Sm$(0.0,1.0) The neutron pairing gap for $^{24}O$ practically vanishes since the first unoccupied orbit 1$d_{3/2}$ is almost 4.5 MeV above the completely filled 2$s_{1/2}$ orbit \cite{Chen2015,Mondal2016}. The pairing correlation energies for a fix gap $\Delta$ is calculated by using the pairing window of 2$\hbar\omega$, where $\hbar\omega$ = $45 A^{-1/3}$ - $25 A^{-2/3}$ MeV \cite{Raj2006,Virender2022b}.
We have obtained three different parametrizations by calibrating the parameters to  a suitable  set of finite nuclei as discussed earlier. Three different parameterizations obtained in the present work give an equally good fit to the properties  of finite nuclei which were used for the optimization procedure.
In Table \ref{tab:table1}, We display the model parameters for the newly generated parameter sets BSRV-CREX, BSRV-PREX and BSRV-CPREX along with theoretical uncertainties/errors. We also list the value of parameters for  NL3   \citep{Lalazissis1997},  FSUGarnet   \citep{Chen2015}, IOPB-1   \citep{Kumar2018} and Big Apple   \citep{Fattoyev2020} for comparison. It can be seen from the table that  comparatively a large value of isovector scalar meson coupling parameter $g_{\delta}$ (4.27816), isovector vector meson coupling $g_{\rho}$ (16.18406)  and  cross-coupling between $\omega$ and $\rho$ meson quantified by the term $\Lambda_{\omega\rho}$ (5.18286) are obtained   for BSRV-CREX parameterization in which only CREX data is included in the fit. The values of coupling parameters $g_{\delta}$, $g_{\rho}$ and $\Lambda_{\omega\rho}$ is more or less the same for BSRV-PREX and BSRV-CPREX models.
In Figs. \ref{par_crex}, \ref{par_prex} and \ref{par_cprex} we show the colour-coded plots for the correlation coefficients between  the coupling parameters appearing in Lagrangian (Eq.\ref{eq:lbm}) for BSRV-CREX, BSRV-PREX and BSRV-CPREX models.
A strong correlation exists between the pairs of coupling parameters $g_{\sigma}$ - $g_{\omega}$ , $\Lambda_{\omega\rho} - g_{\delta}$ , and $\overline{\kappa}-\overline{\lambda}$ for BSRV-CREX model. The strong  correlation is also observed for pair of coupling parameters $g_{\sigma}$ - $g_{\omega}$,  $g_{\rho}$  with $g_{\delta}$ and $\Lambda_{\omega\rho}$. For BSRV-CPREX model, a strong correlation is observed for $g_\delta$ with $g_{\rho}$, $\overline{\lambda}$ and $\zeta$. The coupling parameter $\zeta$ is found to be well correlated with $g_{\rho}$ and $\overline{\lambda}$. A strong correlation between the model parameters indicates a strong interdependence i.e. if one parameter is fixed at a certain value then the other must attain the precise value as suggested by their correlation. It can be seen from Fig. \ref {par_cprex} that when the skin thickness for $^{48}$ Ca and $^{208}$ Pb fitted together there is an overall reduction among the model parameters correlations as compared to those obtained by fitting the individual neutron skin. \\
\subsection{Finite Nuclei and Infinite Nuclear Matter}\label{finite}
\begin{figure}
%\centering
\includegraphics[trim=0 0 0 0,clip,scale=0.5]{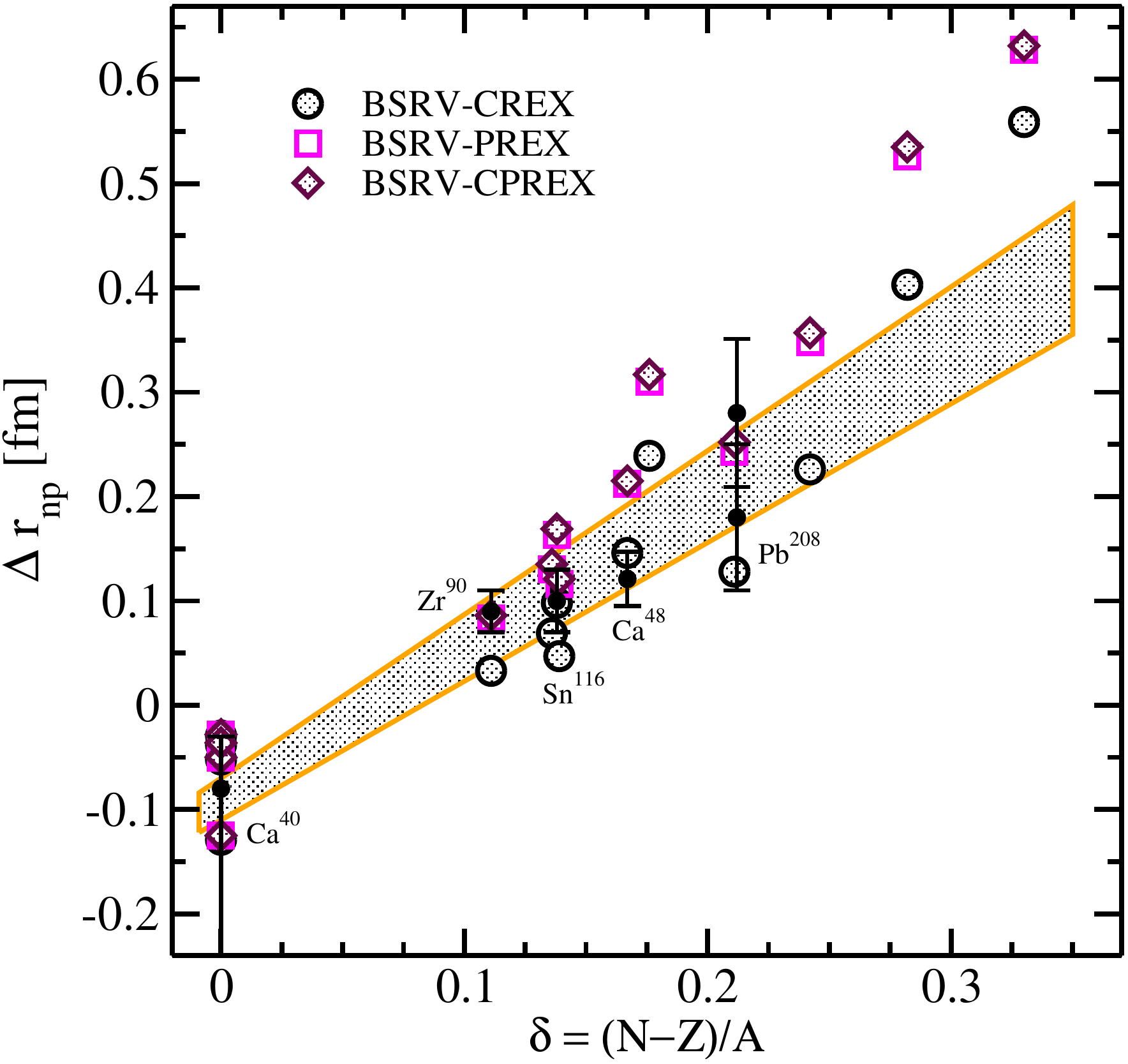}
\caption{\label{nskin} (color online) Variation of neutron skin thickness predicted for the  nuclei considered in the fit data as a function of asymmetry parameter ($\delta$). The shaded region is calculated by using Eq. (\ref{asy}).  }
\end{figure}

\begin{figure}
%\centering
\includegraphics[trim=0 0 0 0,clip,scale=0.5]{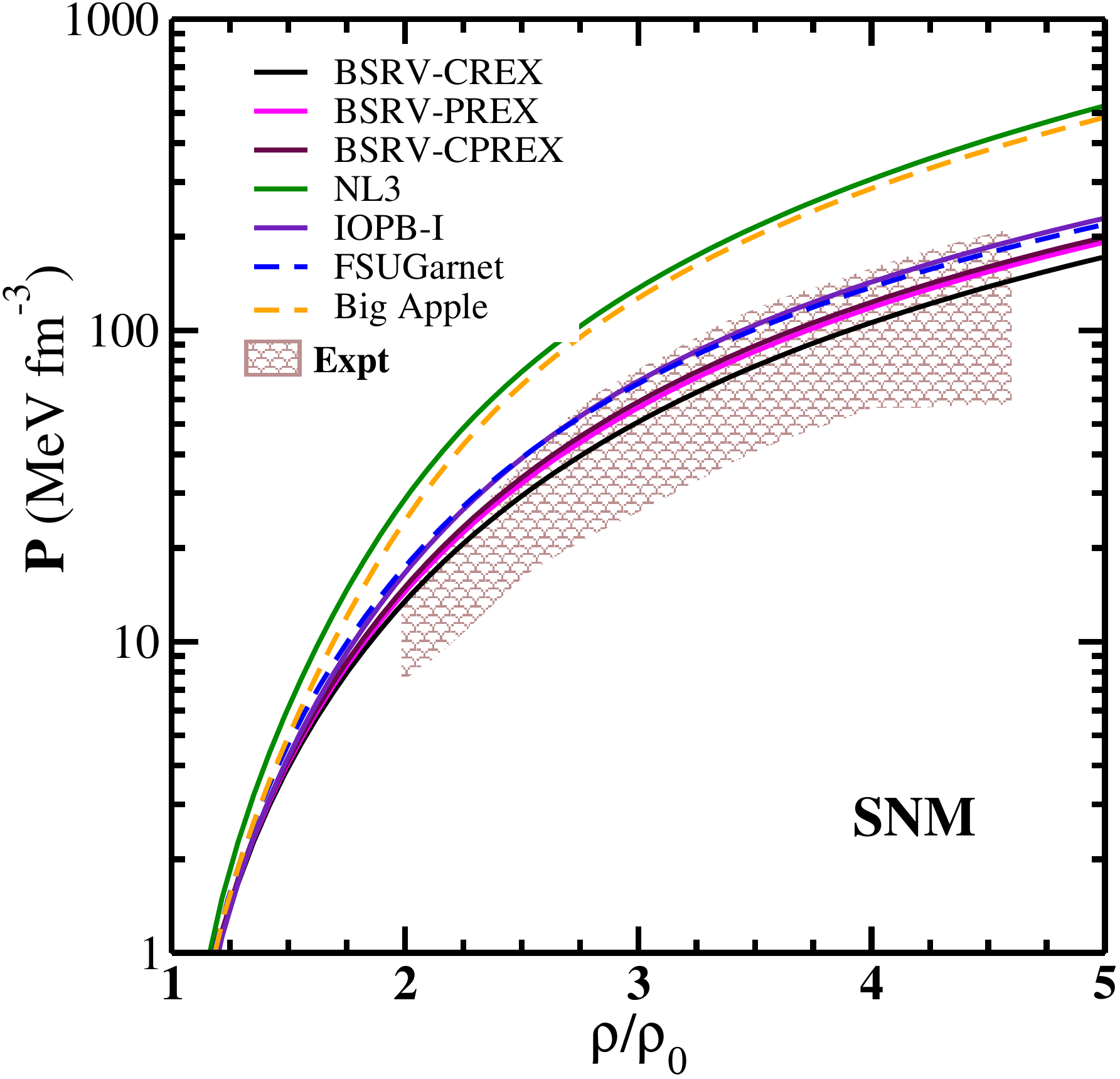}
\caption{\label{snm} (color online) Variation of Pressure as a function of baryon density for symmetric nuclear matter  (SNM) computed with  BSRV's parameterizations along with NL3, IOPB-1, FSUGarnet and Big Apple models. The shaded region represents the experimental data taken from the reference \cite{Danielewicz2002}. }
\end{figure}

\begin{figure}
%\centering
\includegraphics[trim=0 0 0 0,clip,scale=0.5]{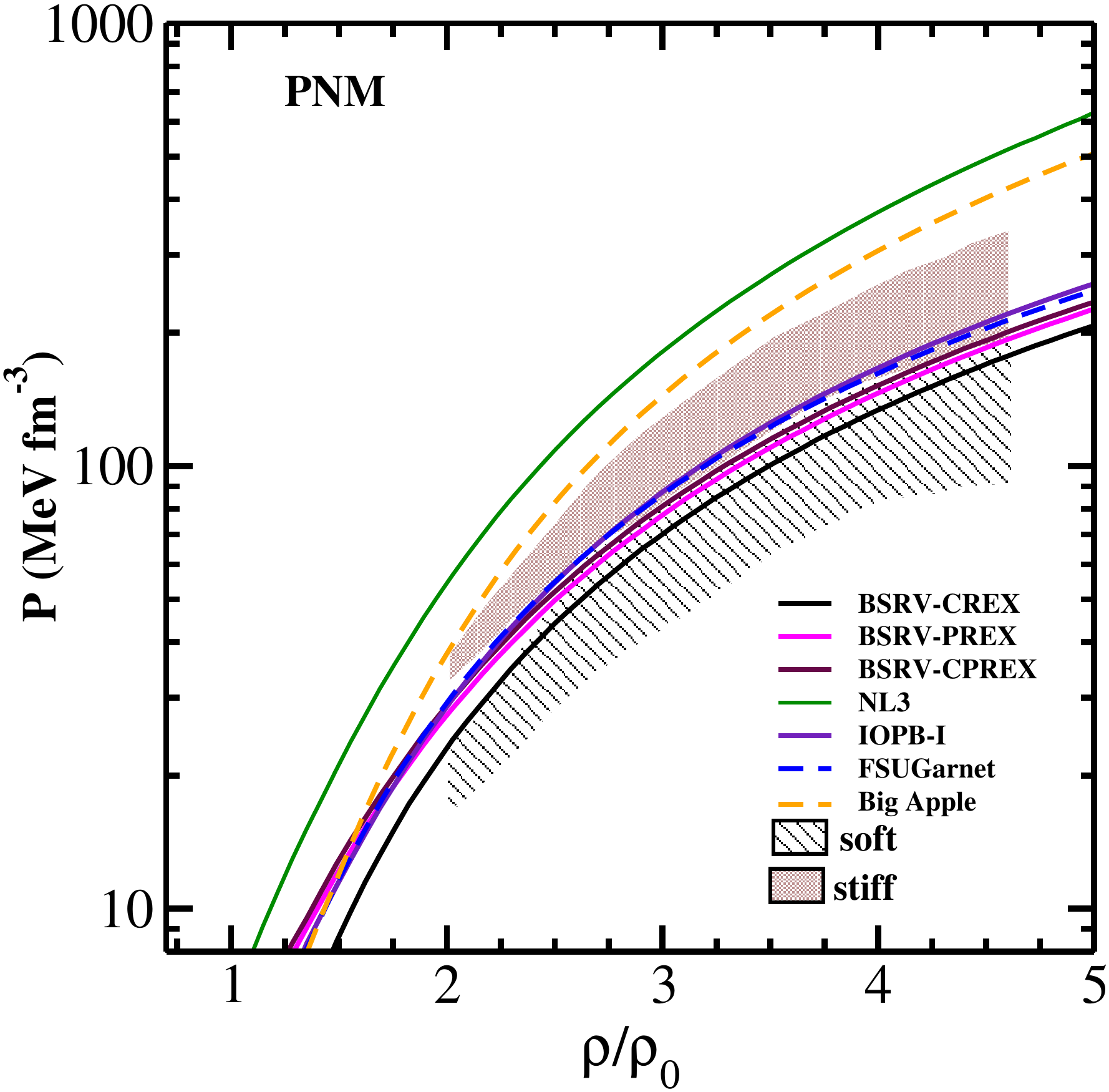}
\caption{\label{pnm} (color online) Variation of Pressure as a function of baryon density for pure neutron matter  (PNM) computed with  BSRV's parameterizations along with NL3, IOPB-1, FSUGarnet and Big Apple models. The shaded region represents the experimental data taken from the reference \cite{Danielewicz2002}. }
\end{figure}

\begin{figure}
%\centering
\includegraphics[trim=0 0 0 0,clip,scale=0.5]{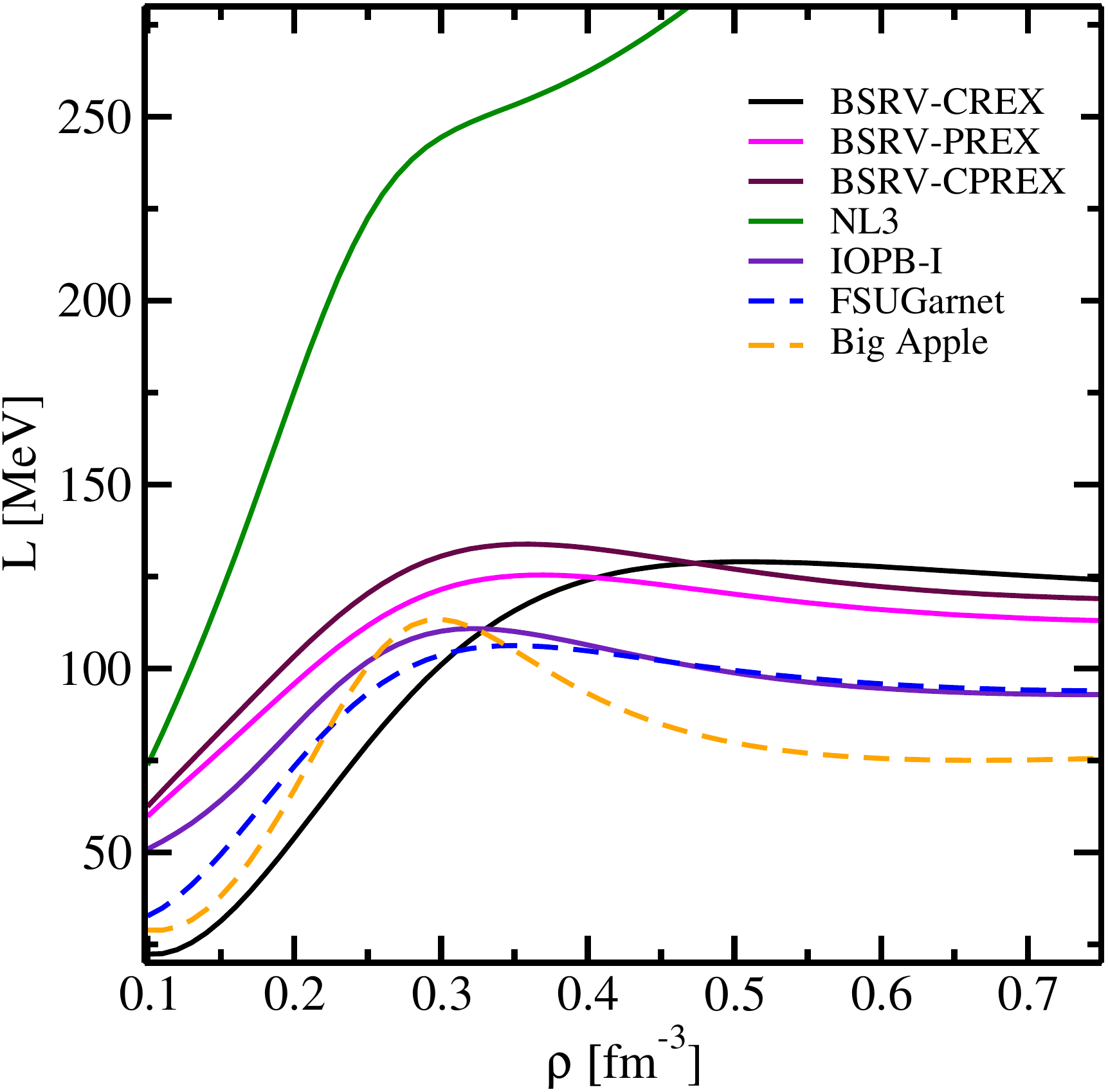}
\caption{\label{lrho} (color online) Variation of density dependence of symmetry energy (L) as a function of baryon density for various parametrizations considered in the present work. }
\end{figure}

In Table \ref{tab:table2}, different observables fitted in the present work, their experimental values \cite{Wang2021,Angeli2013}, adopted errors $\sigma$ on them  along with their calculated values for different BSRV's parametrizations and theoretical uncertainties are displayed. The newly generated parameterizations BSRV-CREX, BSRV-PREX and BSRV-CPREX  give an equally good fit to the properties of finite nuclei. The fitted values of finite nuclei are quite close to their experimental values. The root mean square (rms) errors in total binding energy for all the nuclei considered in our fit are found to be  1.62, 1.39 and  1.39  MeV, whereas the root mean square (rms) errors in total charge radii  are 0.071, 0.080, and  0.081 fm for BSRV-CREX, BSRV-PREX and BSRV-CPREX parameter sets respectively. We also depict  the predicted values of  neutron skin thickness $\Delta r_{np} = R_{n}-R_{p}$  and charge rms radii ($r_{ch}$)for all our parametrizations. It can be observed from the table that for BSRV-CREX parametrization, the value of $\Delta r_{np}$ for $^{48}$Ca nucleus is 0.146 $\pm 0.019$ fm and is consistent with the  recently reported value of $\Delta r_{np}$ of $^{48}$Ca from CREX data \cite{Adhikari2022} while the value of $\Delta r_{np}$ for $^{208}$Pb  nucleus comes out to be 0.13$\pm 0.018$ fm that is also in good agreement with the value reported for neuton skin thickness of $\Delta r_{np} = (0.18\pm 0.07)$ fm  for $^{208}$Pb obtained by dispersive optical model analysis of the Washington University group \cite{Pruitt2020}. The values of $\Delta r_{np}$ obtained for $^{208} Pb$ using the BSRV-CREX model does not satisfy the PREX-II measurement. The values of $\Delta r_{np}$ for $^{208}$Pb obtained for BSRV-PREX and BSRV-CPREX parametrizations is in good agreement with the recently reported value  $\Delta r_{np} = (0.283\pm0.071)$ fm  for $^{208}$Pb for PREX-II data, but the value of $\Delta r_{np}$ predicted for $^{48}$Ca overestimates the CREX data \cite{Adhikari2022}. In Fig \ref{nskin} we show the neutron skin thickness $\Delta r_{np}$ for nuclei considered in our data fit as a function of neutron-proton asymmetry parameter $\delta = \frac{N-Z}{A}$. The predicted result of $\Delta r_{np}$ for the nuclei used in our fit data (Table \ref{tab:table2}) for BSRV-CREX, BSRV-PREX and BSRV-CPREX parametrizations are compared with the corresponding available  experimental values along with error bars taken from Refs.\cite{Jastrzebski2004,Adhikari2021,Adhikari2022,Pruitt2020}.
The shaded regions represent the linear dependence of neutron skin thickness on asymmetry (neutron excess) $\delta$ of a nucleus that can be fitted by \cite{Trzcinska2001,Jastrzebski2004}.
\begin{table*}
\centering
%\begin{table}[p]
\caption{\label{tab:table2}
The calculated values of binding energy (BE) and charge radii ($r_{ch}$)  along with theoretical uncertainties/errors for various BSRV parametrizations are presented. The predicted value of neutron skin thickness $\Delta r _{np}$ =$r_{n}$ -$r_{p}$ is also depicted for  various models. The  corresponding experimental values of BE and $r_{ch}$ \cite{Wang2021,Angeli2013} and $\Delta r_{np}$ \cite{Jastrzebski2004,Adhikari2021,Adhikari2022,Pruitt2020} are also listed. The adopted errors on the observables  ($\sigma$) used for the optimization of parameters and the  asymmetry parameter $\delta = (N-Z)/A$ for the nuclei are  also displayed. The value of BE
are given in units of MeV and  $r_{ch}$, $\Delta r_{np}$ are in  fm.  }
%\begin{ruledtabular}
%\vskip 1cm
\begin{tabular}{|c|c|c|c|c|c|c|c|c|c|}
\hline
\hline
\multicolumn{1}{|c|}{${\bf {Nucleus}}$}&
\multicolumn{1}{|c|}{${\bf {Observables}}$}&
\multicolumn{1}{|c|}{{\bf Exp.}}&
\multicolumn{1}{|c|}{{\bf $\sigma$}}&
%\multicolumn{1}{|c|}{{\bf $\delta$ = $\frac{N-Z}{A}$}}&
\multicolumn{1}{|c|}{{\bf  $\delta$}}&
\multicolumn{1}{|c|}{{\bf BSRV-CREX}}&
\multicolumn{1}{|c|}{{\bf BSRV-PREX}}&
\multicolumn{1}{|c|}{{\bf BSRV-CPREX}}&
\multicolumn{1}{|c|}{{\bf ~~~ NL3}} &
\multicolumn{1}{|c|}{{\bf ~~~~IOPB-I}}
\\
\hline
$^{16}O$ &BE&127.62&4.0&0.0&129.82$\pm$0.32&128.71$\pm$0.55&128.57$\pm$0.60&127.08&128.05\\
 &$r_{ch}$&2.699&0.04&&2.690$\pm$0.024&2.706$\pm$0.031&2.709$\pm$0.029&2.727&2.719\\
  &$\Delta r_{np}$&--&-&&-0.029$\pm$0.020&-0.028$\pm$0.009&-0.028$\pm$0.008&-0.027&-0.029\\
 \hline
$^{24}O$ &BE&168.96&2.0&0.33&166.89$\pm$1.41&170.71$\pm$1.29&171.03$\pm$1.15&170.54&169.48\\
&$r_{ch}$&--&--&&2.753$\pm$0.033&2.734$\pm$0.011&2.732$\pm$0.015&2.737&2.741\\
 &$\Delta r_{np}$&-&-&&0.559$\pm$0.023&0.511$\pm$0.010&0.632$\pm$0.021&0.635&0.613\\
 \hline
 $^{40}Ca$&BE &342.04&3.0&0.0&344.53$\pm$0.47&343.75$\pm$0.56&343.51$\pm$0.81&341.32&342.68\\
 &$r_{ch}$&3.478&0.04&&3.445$\pm$0.035&3.455$\pm$0.029&3.457$\pm$0.014&3.469&3.464\\
   &$\Delta r_{np}$&$-0.08^{+0.05}_{-1.0}$&-&&-0.052$\pm$0.002&-0.050$\pm$0.012&-0.050$\pm$0.002&-0.048&-0.050\\
 \hline
 $^{48}$Ca &BE&415.97&1.0&0.167&416.07$\pm$0.46&415.46$\pm$0.56&415.52$\pm$0.60&414.52&414.57\\
 &$r_{ch}$&3.477&0.04&&3.475$\pm$0.014&3.467$\pm$0.010&3.467$\pm$0.020&3.471&3.471\\
  &$\Delta r_{np}$&0.121$\pm$ 0.026&0.026&&0.146$\pm$0.019&0.212$\pm$0.022&0.215$\pm$0.018&0.226&0.199\\
 \hline
 $^{56}Ni$ &BE&484.01&5.0&0.0&483.92$\pm$0.62&481.55$\pm$1.00&482.14$\pm$0.78&482.12&482.48\\
 &$r_{ch}$&3.750&0.02&&3.709$\pm$0.015&3.721$\pm$0.015&3.718$\pm$0.016&3.716&3.707\\
  &$\Delta r_{np}$&$-0.03^{+0.08}_{-0.11}$&-&&-0.038$\pm$0.001&-0.037$\pm$0.004&-0.036$\pm$0.009&-0.034&-0.037\\
 \hline
 $^{68}Ni$ &BE&590.41&2.0&0.176&592.43$\pm$0.50&592.43$\pm$0.53&592.35$\pm$0.49&591.21&591.66\\
 &$r_{ch}$&--&--&&3.883$\pm$0.015&3.866$\pm$0.027&3.864$\pm$0.020&3.863&3.869\\
  &$\Delta r_{np}$&-&-&&0.239$\pm$0.019&0.310$\pm$0.027&0.317$\pm$0.029&0.333&0.299\\
  \hline
 $^{78}Ni$&BE&642.564 &3.0&0.282&640.19$\pm$1.24&641.61$\pm$1.41&641.83$\pm$1.27&643.04&640.55\\
 &$r_{ch}$&--&--&&3.971$\pm$0.024&3.952$\pm$0.015&3.950$\pm$0.017&3.942&3.950\\
  &$\Delta r_{np}$&-&-&&0.403$\pm$0.033&0.526$\pm$0.123&0.535$\pm$0.034&0.553&0.506\\
 \hline
 $^{88}Sr$ &BE&768.42&2.0&0.136&768.51$\pm$0.51&767.69$\pm$0.55&767.62$\pm$0.56&767.31&766.65\\
 &$r_{ch}$&&0.02&&4.237$\pm$0.013&4.227$\pm$0.012&4.226$\pm$0.014&4.225&4.229\\
  &$\Delta r_{np}$&-&-&&0.069$\pm$0.009&0.130$\pm$0.026&0.134$\pm$0.021&0.148&0.118\\
 \hline
 $^{90}Zr$ &BE&783.81&1.0&0.111&783.92$\pm$0.59&783.19$\pm$0.60&783.16$\pm$0.62&782.95&782.44\\
 &$r_{ch}$&4.269&0.02&&4.289$\pm$0.018&4.282$\pm$0.012&4.282$\pm$0.014&4.280&4.284\\
  &$\Delta r_{np}$&$0.09^{+0.02}_{-0.02}$&-&&0.033$\pm$0.013&0.083$\pm$0.022&0.086$\pm$0.018&0.097&0.072\\
 \hline
$^{100}Sn$&BE&825.10&2.0&0.0&827.55$\pm$0.97&826.96$\pm$1.04&827.55$\pm$1.34&829.33&827.52\\
 &$r_{ch}$&--&--&&4.515$\pm$0.014&4.522$\pm$0.014&4.521$\pm$0.026&4.511&4.514\\
  &$\Delta r_{np}$&-&-&&-0.129$\pm$0.002&-0.126$\pm$0.018&-0.125$\pm$0.002&-0.117&-0.126\\
   \hline
 $^{116}Sn$ &BE&988.67&2.0&0.138&987.42$\pm$0.65&987.15$\pm$0.72&986.85$\pm$0.69&986.89&986.24\\
 &$r_{ch}$&4.627&0.02&&4.631$\pm$0.017&4.617$\pm$0.015&4.617$\pm$0.014&4.610&4.620\\
  &$\Delta r_{np}$&$0.10^{+0.03}_{-0.03}$&-&&0.098$\pm$0.013&0.163$\pm$0.024&0.169$\pm$0.028&0.183&0.149\\
 \hline
 $^{132}Sn$&BE&1102.22 &2.0&0.242&1102.55$\pm$0.94&1102.51$\pm$1.05&1102.37$\pm$0.96&1104.81&1101.93\\
 &$r_{ch}$&4.709&0.02&&4.742$\pm$0.013&4.723$\pm$0.019&4.722$\pm$0.014&4.710&4.721\\
  &$\Delta r_{np}$&-&-&&0.226$\pm$0.025&0.348$\pm$0.048&0.357$\pm$0.025&0.383&0.325\\
 \hline
 $^{144}Sm$&BE&1195.77&2.0&0.139&1197.69$\pm$0.82&1197.21$\pm$0.91&1196.90$\pm$0.82&1198.14&1195.82\\
 &$r_{ch}$&--&--&&4.978$\pm$0.021&4.965$\pm$0.102&4.965$\pm$0.014&4.956&4.967\\
  &$\Delta r_{np}$&-&-&&0.047$\pm$0.013&0.116$\pm$0.023&0.121$\pm$0.031&0.137&0.103\\
 \hline
 $^{208}$Pb &BE&1636.34&1.0&0.212&1637.11$\pm$0.92&1636.57$\pm$0.94&1636.00$\pm$0.93&1639.43&1636.75\\
 &$r_{ch}$&5.501&0.04&&5.551$\pm$0.016&5.533$\pm$0.018&5.532$\pm$0.029&5.517&5.532\\
 &$\Delta r_{np}$&$0.283\pm0.071$&0.071&&0.130$\pm$0.018&0.243$\pm$0.048&0.252$\pm$0.028&0.279&0.219\\
 &&$(0.18\pm0.07) $&-&&&&&&\\
\hline
\hline
\end{tabular}
%\end{ruledtabular}
%\end{table}
\end{table*}
\begin{equation}
\label{asy}
\Delta r _{np}= (-0.09\pm 0.02)+(1.45\pm 0.12)\delta ~~~~~   (fm)
\end{equation}
The values of $\Delta r_{np}$ obtained for BSRV-PREX  and BSRV-CPREX parameter sets for some of the nuclei deviate from shaded region as can be observed from the Fig. \ref{nskin}. This may be attributed to the fact that for these paramerizations the value of   $\omega$ -$\rho$ meson coupling parameter $\Lambda_{\omega\rho}$ that constrains the density dependence of symmetry energy and coupling parameters $g_{\delta}$ and $g_{\rho}$ is relatively smaller as compared to BSRV-CREX parameter set. For BSRV-CREX parameter set, the values of $\Delta r_{np}$ lie in the shaded region or very close to it except for $^{24}O$ nucleus for which asymmetry is 0.33.
For BSRV-CREX, BSRV-PREX and BSRV-CPREX models, the calculated values of neutron skin thickness for $^{208}$Pb nucleus are 0.130$\pm0.018$ fm, 0.243$\pm0.048$ fm and 0.252$\pm0.028$ fm and  for $^{48}$Ca nucleus are 0.146$\pm$0.019 fm, 0.212$\pm0.022$ fm and 0.215$\pm$0.018 fm respectively. The value of  $\Delta r_{np}$=0.146$\pm$0.019 fm  for $^{48}$Ca predicted by BSRV-CREX is consistent with the recently measured $\Delta r_{np} =0.121\pm0.026$ fm from parity-violating electron scattering experiment (CREX) \cite{Adhikari2022}.  The values of neutron skin thickness $\Delta r_{np}$= 0.243$\pm0.048$ fm and 0.252$\pm0.028$ fm predicted for $^{208}$Pb by BSRV-PREX  and BSRV-CPREX are also in agreement  with the recently reported neutron skin thickness from updated Lead Radius Experiment(PREX-II) \cite{Adhikari2021}.\\
In the present work, we have tried to resolve the serious conflict of CREX and PREX-II measurements by searching the model parameters of BSRV-CPREX parameterization by including  both the $\Delta r_{np}$ of $^{48}$Ca  from the CREX  and $^{208}$Pb  from PREX-II  in our fit data for the optimization of model parameters. But the $\Delta r_{np}$ predicted by this model overestimates the recently reported value by  CREX  and the serious conflict between CREX and PREX-II measurements  continues as also discussed in Ref. \cite{Par2022}.\\
In Table \ref{tab:table3}, we present our results for the symmetric nuclear matter (SNM) properties such as binding energy per nucleon (E/A), incompressibility (K) ,  symmetry energy coefficient (J), density dependence of symmetry energy (L) and ratio of effective mass to the mass of nucleon at the saturation density ($\rho_{0}$) and curvature of symmetry energy ($K_{sym}$) along with theoretical uncertainties. These properties play a vital role for constructing the EoS for nuclear matter. E/A is more or less the same  for all BSRV's parameterizations. For  newly generated parameterizations BSRV-PREX and BSRV-CPREX,  the  value of J and L are consistent with the constraints from observational analysis J = 38.1 $\pm$ 4.7 MeV  and L = 106 $\pm$ 37 MeV  as reported by Reed et. al.,\cite{Reed2021} and for  BSRV-CREX parameter set, the  value of J = 28.97$\pm$0.99 MeV and L = 30.61$\pm$ 6.74 MeV is in close proximity to that reported in Ref.\cite{Par2022}, it is also consistent with the constraints from observational analysis J = 31.6 $\pm$ 2.66 MeV \citep{Li2013}. The neutron skin thickness as reported by CREX collaboration suggests softness i.e. low value of symmetry energy coefficient (J) and its corresponding density dependence (L). The value of K lies in the range 222.29$\pm$13.08 - 227.45$\pm$6.95 MeV which is also in good agreement with the value of K = 240 $\pm$ 20 MeV determined from isoscalar giant monopole resonance (ISGMR) for  $^{90}Zr$ and $^{208}$Pb nuclei  \citep{Colo2014,Piekarewicz2014}. The curvature of symmetry energy $K_{sym}$ also satisfies the empirical limit discussed in \cite{Zimmerman2020}. The  ratio of effective mass to the nucleon mass is found to be similar for all BSRV's parameterizations as shown in Table \ref{tab:table3}. The SNM properties calculated with NL3, FSUGarnet, IOPB-1 and Big Apple are also shown for comparison.\\
In Fig. (\ref{snm} and \ref{pnm}), we plot the EoS i.e. pressure as a function of baryon density scaled to saturation density ($\frac{\rho}{\rho_{0}}$) for SNM and pure neutron matter (PNM) using BSRV-CREX, BSRV-PREX and BSRV-CPREX parameterizations which are in good agreement and lie in the allowed region  with the EoS extracted from the analysis of particle flow in heavy ion collision   \citep{Danielewicz2002}.
 These results are also compared with the NL3, IOPB-1, FSUGarnet and Big Apple parameterizations. It can be easily seen that the EoSs for SNM and PNM obtained from  NL3 and Big Apple parameterizations are very stiff and are ruled out by constraints imposed by heavy ion collision data. The stiffness of the EoSs for NL3 and Big Apple  parameter sets may be due to the fact that the  coupling parameter $\zeta$  which  is responsible for varying  the high-density behavior of EoS is zero for NL3 and very small for Big Apple parameter sets.
 The EoSs calculated using BSRV's parameter sets are relatively much softer  and lie in the allowed region of  heavy ion collision data   \citep{Danielewicz2002}. The EoS calculated from BSRV-CREX parametrization is  the softest amongst all EoSs and it might  be due to the relatively somewhat  higher value of parameter $\zeta$ and $\Lambda_{\omega\rho}$ obtained for this parameter set during the calibration procedure which is responsible for varying high-density behavior of EoS. In Fig.\ref{lrho} we plot the density dependence of symmetry energy (L) as a function of baryon density for BSRV's paramerizations. The results  for NL3,IOPB-1, FSUGarnet and Big Apple parameter sets are also displayed for comparison. It can be seen from the figure that in low or medium density regime the behvior of BSRV-CREX is softest (Low value of L at a given baryon density) amongst all parametrizations and changes to stiffest in high density regime even though the $\Delta r_{np}$ for $^{208}$Pb nucleus is smallest for this parameter set. This may be attributed to the large value of coupling parameter $g_{\delta}$  (4.27816) obtained for BSRV-CREX parameter set. The coupling parameter $g_{\delta}$ is responsible for changing the behivior of L from soft in low - medium density regime to stiff in high density regime. The stiffness of L for BSRV-PREX and BSRV-CPREX parameter sets may also be due to the coupling parameter $g_{\delta}$ and large value of $\Delta r_{np}$ for $^{208}$Pb nucleus.  A large value of $\Delta r_{np}$ for $^{208}$Pb NL3 parameter  and  its small value for  Big Apple may be responsible for for stiffness and softness  behaviour of L respectively.
 \\
\begin{table*}
\centering
%\begin{table}[p]
\caption{\label{tab:table3}
The bulk nuclear matter properties at saturation density along with theoretical uncertainties  for BSRV-CREX, BSRV-PREX and BSRV-CPREX  parametrizations are listed along with NL3, IOPB-1, FSUGarnet and Big Apple models.
$\rho_{0}$, E/A, K, J, L, $K_{sym}$ and $ M^{*}/M$ denotes the saturation density, binding energy
per nucleon, incompressibility coefficient, symmetry energy, density dependence of symmetry energy, the curvature of symmetry energy  and the ratio of effective nucleon mass to the nucleon mass, respectively.}
%\begin{ruledtabular}
%\vskip 1cm
\begin{tabular}{cccccccc}
\hline
\hline
\multicolumn{1}{c}{${\bf {Parameters}}$}&
\multicolumn{1}{c}{{\bf BSRV-CREX}}&
\multicolumn{1}{c}{{\bf BSRV-PREX}}&
\multicolumn{1}{c}{{\bf BSRV-CPREX}}&
\multicolumn{1}{c}{{\bf NL3}}&
\multicolumn{1}{c}{{\bf IOPB-1}}&
\multicolumn{1}{c}{{\bf FSUGarnet}}&
\multicolumn{1}{c}{{\bf BigApple}}\\
\hline
${\bf{\rho_{0} ~(fm^{-3})}}$&0.148$\pm$0.003&0.148$\pm$0.001&0.148$\pm$0.002&0.148&0.149&0.153&0.155\\
${\bf E/A ~(MeV)}$ &-15.99$\pm$0.03&-16.10$\pm$0.06&-16.09$\pm$0.05&-16.24&-16.09&-16.23&-16.34\\
${\bf K ~(MeV)}$ &222.29$\pm$13.08&227.45$\pm$6.95&226.99$\pm$3.74&271.56&222.57&229.62&227.09\\
${\bf ~M^{*}/M}$&0.600$\pm$0.007&0.606$\pm$0.004&0.602$\pm$0.006&0.595&0.593&0.578&0.608\\
${\bf J ~(MeV)}$&28.97$\pm$0.99&34.41$\pm$2.71&34.99$\pm$2.15&37.40&33.30&30.98&31.41\\
${\bf L ~(MeV)}$&30.61$\pm$6.74&77.08$\pm$28.87&82.32$\pm$22.93&118.56&63.85&50.92&40.33\\
${\bf ~K_{sym}(MeV)}$&61.79$\pm$32.74&-71.48$\pm$5.53&-65.65$\pm$20.32&100.90&-37.79&58.46&89.58\\
\hline
\hline
\end{tabular}
%\end{ruledtabular}
%\end{table}
\end{table*}


\subsection{Neutron star properties}
%\subsubsection{Equatin of State and Mass-Radius Relationship}
\begin{figure}
\centering
\includegraphics[trim=0 0 0 0,clip,scale=0.5]{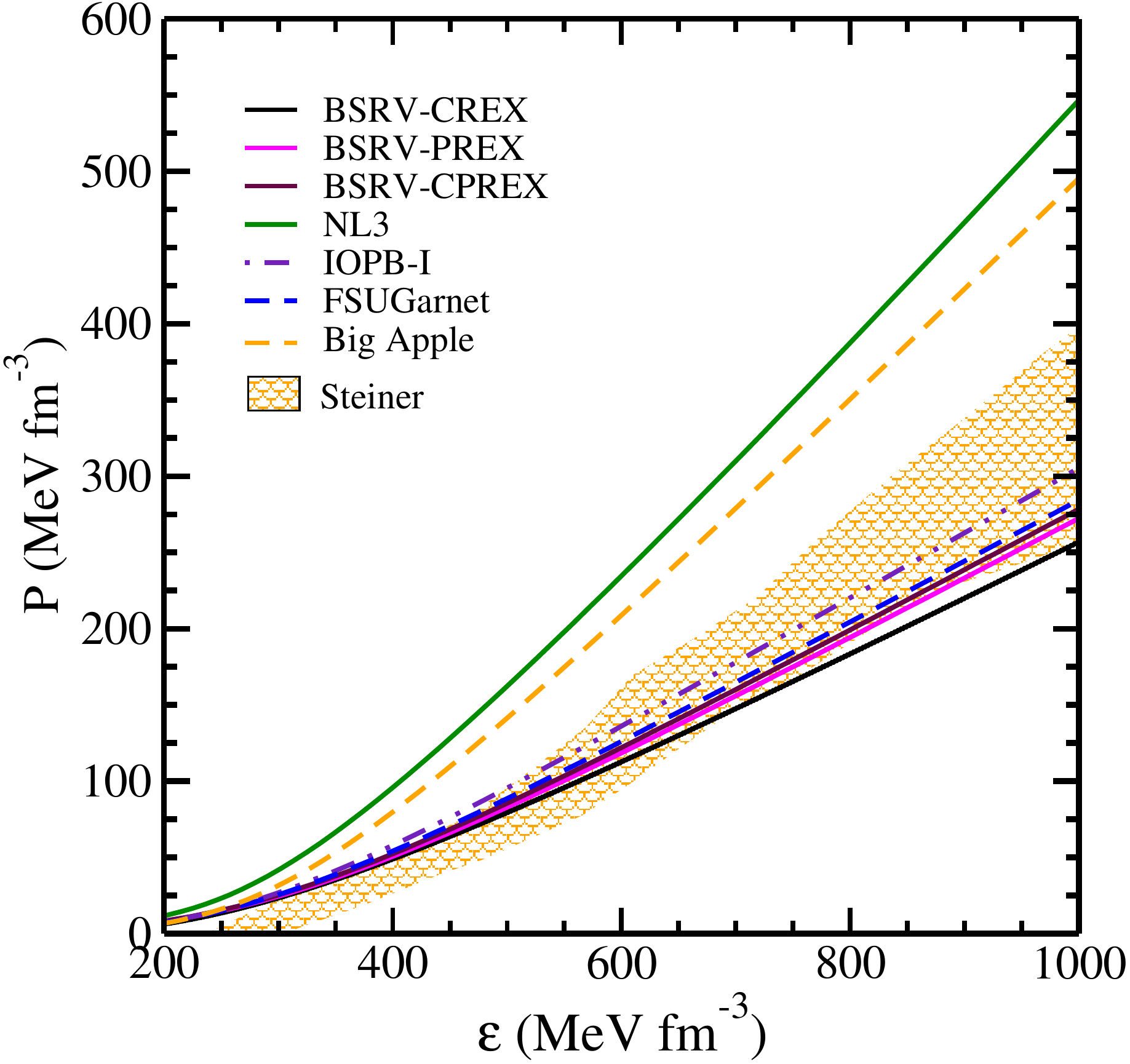}
\caption{\label{eos} (Color online) EoSs i.e. pressure as a function of energy density for $\beta$-stable equilibrated nuclear matter for various parametrizations.The  shaded region represents the observational constraints reported in Ref. \cite{Steiner2010}.}
\end{figure}
In Fig (\ref{eos}) we display the EoS i.e.  the variation of  pressure with the energy density  for the neutron star in $\beta$ equilibrium for  all  BSRV's  parameterizations. The results calculated with  NL3, IOPB-1, FSUGarnet and Big Apple parameter sets are also shown for comparison. The shaded region in Fig. \ref{eos} represents the observational constraints at $r_{ph}$=R with the uncertainty of 2$\sigma$ \cite{Steiner2010}. Here $r_{ph}$ and R  are the photospheric and neutron star radius respectively. It can be observed from the figure  that the EoS computed with  newly generated  BSRV's parameter sets are softer and lie in the lower boundary of the shaded region at very high density ($\cal E$ $\approx$ 700-1000 MeV fm$^{-3}$). The softness of EoSs for BSRV's parameter sets may be due to the moderate value of $\omega$ meson coupling parameter $\zeta$ that governs the high-density behavior of EoS. The EoSs calculated for NL3 and Big Apple parameter sets are stiffer and may be attributed to either zero or very  small value of coupling parameter $\zeta$ for these parameter sets respectively.
\begin{figure}
\centering
\includegraphics[trim=0 0 0 0,clip,scale=0.42]{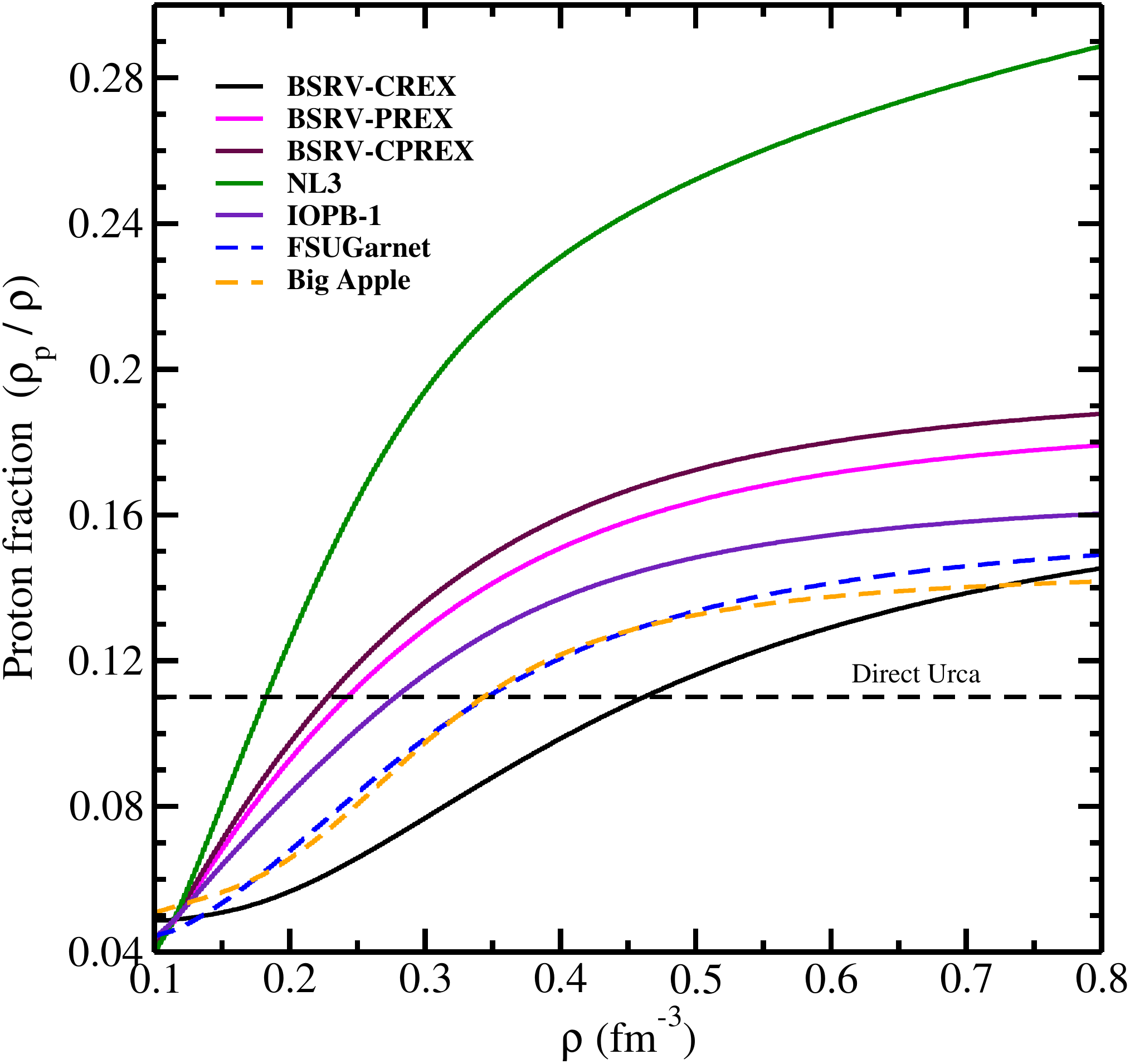}
\caption{\label{pfraction} (Color online) The plot of proton fractions as a function of baryon density. The dotted horizontal line represents the threshold for the direct Urca process i.e. $(\rho_{p}/\rho)$ = 0.11 \cite{Maruyama1999}.}
\end{figure}
In Fig. \ref{pfraction} we illustrate the proton fraction in neutron star matter. One remarkable point is that the $g_{\delta}$ has an influence on proton fraction at high densities. The large value of $g_{\delta}$ for BSRV-CREX suppresses the proton fraction and then delays the direct Urca process in which neutrinos can be emitted rapidly. It can also be observed that $\Delta r_{np}$ of  $^{208}$Pb nucleus also plays  a significant role in the direct Urca process. The large value of $\Delta r_{np}$  for $^{208}$Pb may be attributed to the early start of the direct Urca process for BSRV-PREX, BSRV-CPREX and NL3 whereas its small value for BSRV-CREX parameter set may be responsible for delayed  direct Urca process for BSRV-CREX parameterization.\\
\begin{table*}
 \caption{\label{tab:table4}The properties of nonrotating  neutron stars for the various EoSs computed with  BSRV's parameter sets are displayed along with theoretical uncertainties on them. M$_{max}(M_\odot)$ and  R$_{max}$  denote the Maximum Gravitational mass and  radius corresponding to the maximum mass  of the nonrotating neutron  stars respectively. The values for R$_{max}$, R$_{1.4}$, and  $\Lambda_{1.4}$ denote radius and  dimensionless tidal deformability  corresponding to $M_{max}$ and 1.4M$_\odot$.}
 \begin{tabular}{cccccccc}
\hline
\bf{SN}&\bf{EoS}& \bf{M$_{max}$(M$_{\odot}$)}& \bf{~~R$_{max}$ } & \bf{~R$_{1.4}$}&${\bf{~~\Lambda_{1.4}}}$\\
 & & & (km)& ~~(km) ~~ ~~~&&\\
 \hline
 1.& BSRV-CREX&1.95$\pm$0.04&11.34$\pm$0.21&12.66$\pm$0.39&525.84$\pm$151.48\\
2& BSRV-PREX&2.01$\pm$0.04&11.67$\pm$0.26&13.27$\pm$0.55&638.51$\pm$142.03\\
3.& BSRV-CPREX&2.04$\pm$0.04&11.79$\pm$0.29&13.41$\pm$0.71&682.57$\pm$219.73\\
4.& NL3&2.77&13.28&14.65&1274\\
5& IOPB-I&2.15&11.95&13.29&682\\
6& FSUGarnet&2.06&11.65&12.98&622.51\\
7& Big Apple&2.60&12.41&13.12&715.96\\
 \hline
\hline
\end{tabular}
\end{table*}
\begin{figure}
\centering
\includegraphics[trim=0 0 0 0,clip,scale=0.5]{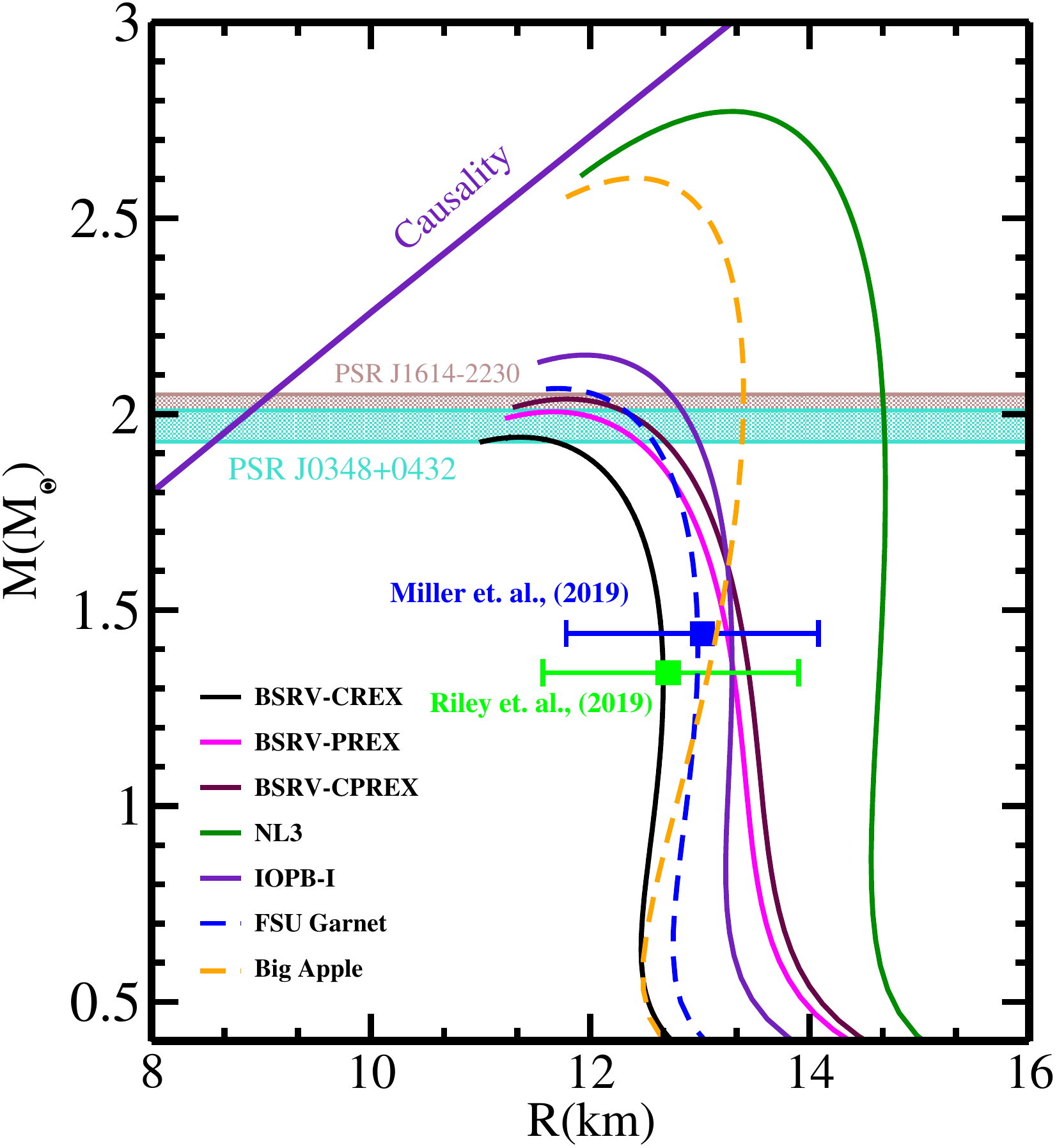}
\caption{\label{mr} (Color online) Variation of  the gravitational mass of non-rotating neutron stars as a function of radius.
 for BSRV's parameterizations. The results for NL3,IOPB-1, FSUGarnet and Big Apple parameter sets are also shown.}
\end{figure}
The mass and radius of a neutron star are obtained by solving the Tolman-Oppenheimer-Volkoff (TOV) equations \cite{Oppenheimer1939,Tolman1939} given as:
\begin{equation}
\label{eq:tov}
\frac{dP(r)}{dr} = -\frac{\{\epsilon(r)+P(r)\}\{4\pi r^3 P(r)+m(r)\}}{r^2(1-2m(r)/r)}
\end{equation}
\begin{equation}
\label{eq:nr31}
\frac{dm}{dr}=4\pi r^2\epsilon(r),
\end{equation}
\begin{equation}
m(r)= 4\pi\int_0^{r}dr r^2 \epsilon(r)
\end{equation}
where $P(r)$  is the pressure at radial distance $r$ and $m(r)$  is the  mass of neutron stars  enclosed in the sphere of radius $r$.
In Fig. \ref{mr} we plot the results for  the gravitational mass of a non-rotating neutron  star and its radius  for BSRV-CREX, BSRV-PREX and BSRV-CPREX parameter sets. The results are also displayed for  NL3, IOPB-1, FSUGarnet and Big Apple parameterizations. It is observed that the maximum gravitational mass of the non-rotating neutron star for BSRV's parameter sets lies in the range 1.95$\pm$0.04M$_{\odot}$  - 2.04$\pm$0.04 M$_{\odot}$  which is in good agreement with the mass constraints from GW170817 event, pulsars PSRJ1614-2230, PSRJ0348+0432, and PSRJ0740+6620  \cite{Demorest2010,Fonseca2016,Rezzolla2018,Riley2021,Miller2021}.
The EoS for $\beta$- equalibrated matter calculated with BSRV-CREX model is the softest amongst all parametrizations considered in this work. The radius  of canonical mass ($R_{1.4}$) including BPS crust \cite{Baym1971} for low density region is 12.66 $\pm$ 0.39 Km, 13.27$\pm$0.55 Km and 13.41$\pm0.71$ Km for BSRV-CREX, BSRV-PREX and BSRV-CPREX parameterizations  which  satisfies the radius constraints from NICER on $R_{1.4}$ and PREX-II  reported in  Ref. \cite{Reed2021}. The value of  $R_{1.4}$ for BSRV's paramerization is consistent with the softness and stiffness behaviour of symmetry energy coefficient. The value of  $R_{1.4}$ for NL3 parameterization is 14.65 Km  which seems to be ruled out by the constraints for $R_{1.4}$ extracted from Ref. \cite{Annala2017}.\\
%\subsubsection{Tidal deformability}
The tidal deformability rendered by the companion stars on each other in a binary system can provide significant information on the EoS of neutron stars \cite{Hinderer2008,Hinderer2010}.
The tidal influences of its companion in the binary neutron star (BNS) system will deform neutron stars in the binary system and, the resulting change in the gravitational potential modifies 
the BNS orbital motion and its corresponding gravitational wave (GW) signal. This effect on GW phasing can be parameterized by the dimensionless tidal deformability parameter,
$\Lambda_i = \lambda_i/M_i^5,$ i = 1, 2.  For each neutron star, its quadrupole moment ${\cal{Q}}_{j,k}$ must be related to the tidal field ${\cal{E}}_{j,k}$ caused by its companion
as, ${\cal{Q}}_{j,k} = -\lambda {\cal {E}}_{j,k}$, where, $j$ and $k$ are spatial tensor indices.
The dimensionless tidal deformability
parameter  $\Lambda$  of a static, spherically symmetric compact star
depends on the neutron star compactness parameter C and a dimensionless quadrupole Love number k$_{2}$ as, $\Lambda$=(2k$_{2}/{3)C^{-5}}$. The $\Lambda$ critically parameterizes 
the deformation of neutron stars under the given tidal field, therefore it should depend on the EoS of nuclear dense matter.
To measure the Love number k$_{2}$ along with the evaluation of the TOV  
equations we have to compute y$_{2}$ = y(R) with 
initial boundary condition y(0) = 2 from the first-order
differential equation   \citep{Hinderer2008,Hinderer2009,Hinderer2010,Damour2010} simultaneously, 
 \begin{eqnarray}
 \label{y}
   y^{\prime}=\frac{1}{r}[-r^2Q-ye^{\lambda}\{1+4\pi Gr^2(P-{\cal{E}})\}-y^{2}], 
 \end{eqnarray}
where  Q $\equiv$ 4$\pi$Ge$^{\lambda}$(5${\cal{E}}$+9P+$\frac{{\cal{E}}+P}{c_{s}^2})$
 -6$\frac{e^{\lambda}}{r^2}$-$\nu^{\prime^2}$ and       
  e$^{\lambda} \equiv(1-\frac{2 G m}{r})^{-1}$ and, $\nu^{\prime}\equiv$ 2G e$^{\lambda}$ 
       ($\frac{m+4 \pi P r^3}{r^2}$).
First, we get the solutions of Eq.(\ref{y}) with boundary condition, y$_{2}$ = y(R),
then the electric tidal Love
number k$_{2}$ is calculated from the expression as,
\begin{eqnarray}
 k_{2}=\frac{8}{5}C^{5}(1-2C)^{2}[2C(y_{2}-1)-y_{2}+2]\{2C(4(y_{2}+1)C^4\nonumber\\
 +(6y_{2}-4)C^{3}
 +(26-22y_{2})C^2+3(5y_{2}-8)C-3y_{2}+6)\nonumber\\
 -3(1-2C)^2(2C(y{_2}-1)-y_{2}+2) \log(\frac{1}{1-2C})\}^{-1}.\nonumber\\
\end{eqnarray}
\begin{figure}
\centering
\includegraphics[trim=0 0 0 0,clip,scale=0.5]{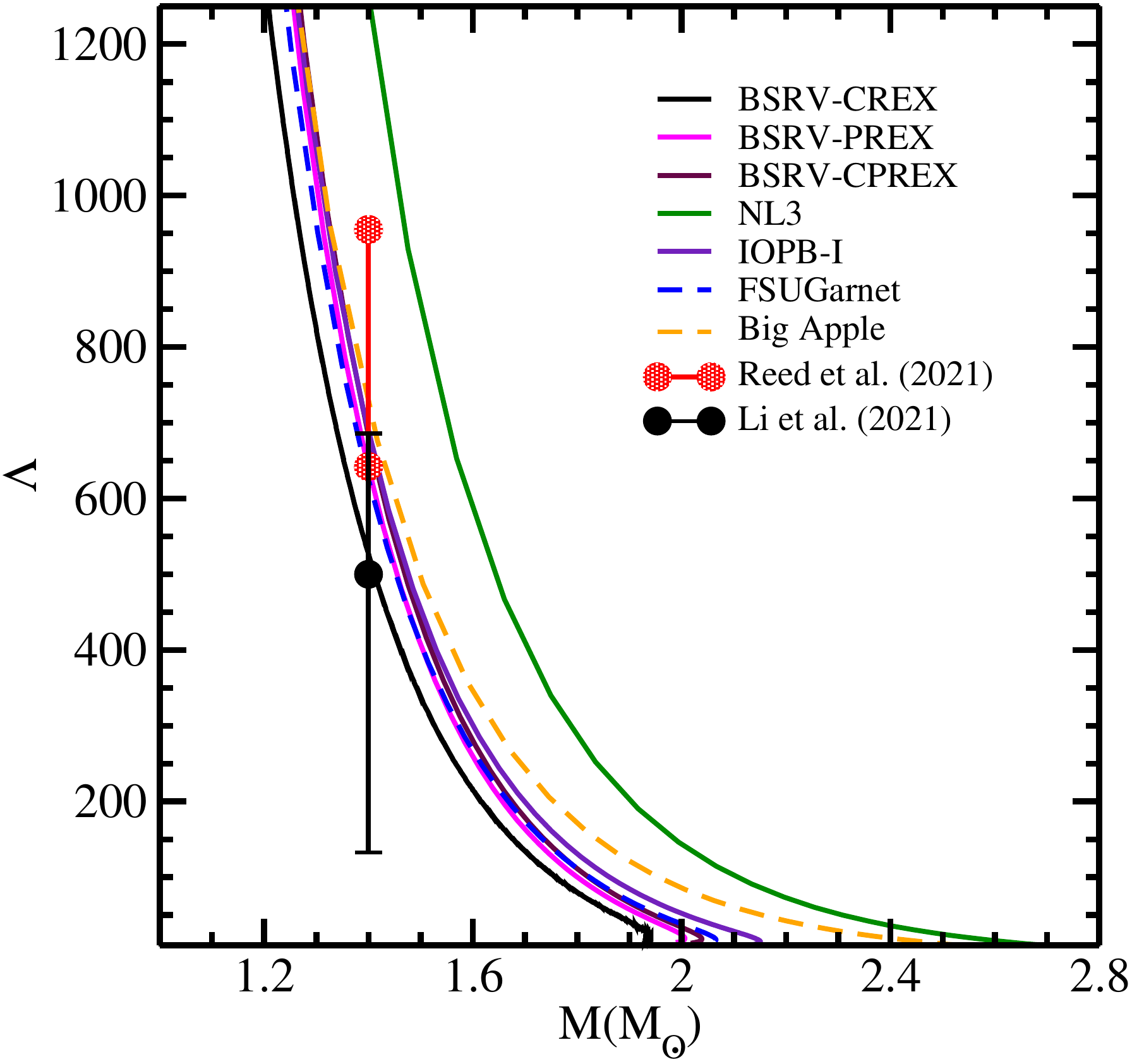}
\caption{\label{tidal} (Color online) Variation of  dimensionless tidal deformability ($\Lambda$)
 with respect
to gravitational mass for all BSRV's  parameterizations.  The results for NL3, IOPB-1, FSUGarnet and Big Apple  parameter sets  are also shown.}
\end{figure}
In Fig. (\ref{tidal}), we display  the results of dimensionless tidal deformability $\Lambda$ as a function of gravitational mass for neutron stars for BSRV's parametrizations. For the sake of comparison, we also display the results calculated with NL3,  IOPB-1, FSUGarnet and Big Apple parameterizations. The values of $\Lambda_{1.4}$ obtained for canonical mass are  525.84$\pm$151.48, 638.51$\pm$142.03 and 682.57$\pm$219.73 corresponding to  BSRV-CREX, BSRV-PREX and BSRV-CPREX respectively and satisfy the constraints as reported in Ref. \cite{Reed2021,Chen2021,Abbott2017}.  It may be noted that the value of  tidal deformability of 1.4$M_{\odot}$ neutron star obtained for BSRV-PREX and BSRV-CPREX  models have overlap with the  revised limit of $\Lambda_{1.4}\le 580$ within 1$\sigma$ uncertainty \cite{Abbott2018}. The value (525.84$\pm$151.48) of $\Lambda_{1.4}$ obtained for the BSRV-CREX parameter set is also consistent with the constraints imposed in Ref. \cite{Abbott2018}. This  might be  attributed to the  high value of coupling  $g_{\delta}$ for this parametrization that governs the  softness and stiffness of EoS in the low and the high-density regime.\\
In Table \ref{tab:table4}, we summarize the results for non-rotating neutron star properties such as maximum gravitational mass (M), neutron star radius corresponding to the maximum mass ($R_{max}$), radius $R_{1.4}$, radius $R_{2.0}$ and tidal deformability ($\Lambda$) corresponding to canonical and maximum mass of neutron star along with theoretical uncertainties.
\begin{figure}
\centering
\includegraphics[trim=0 0 0 0,clip,scale=0.5]{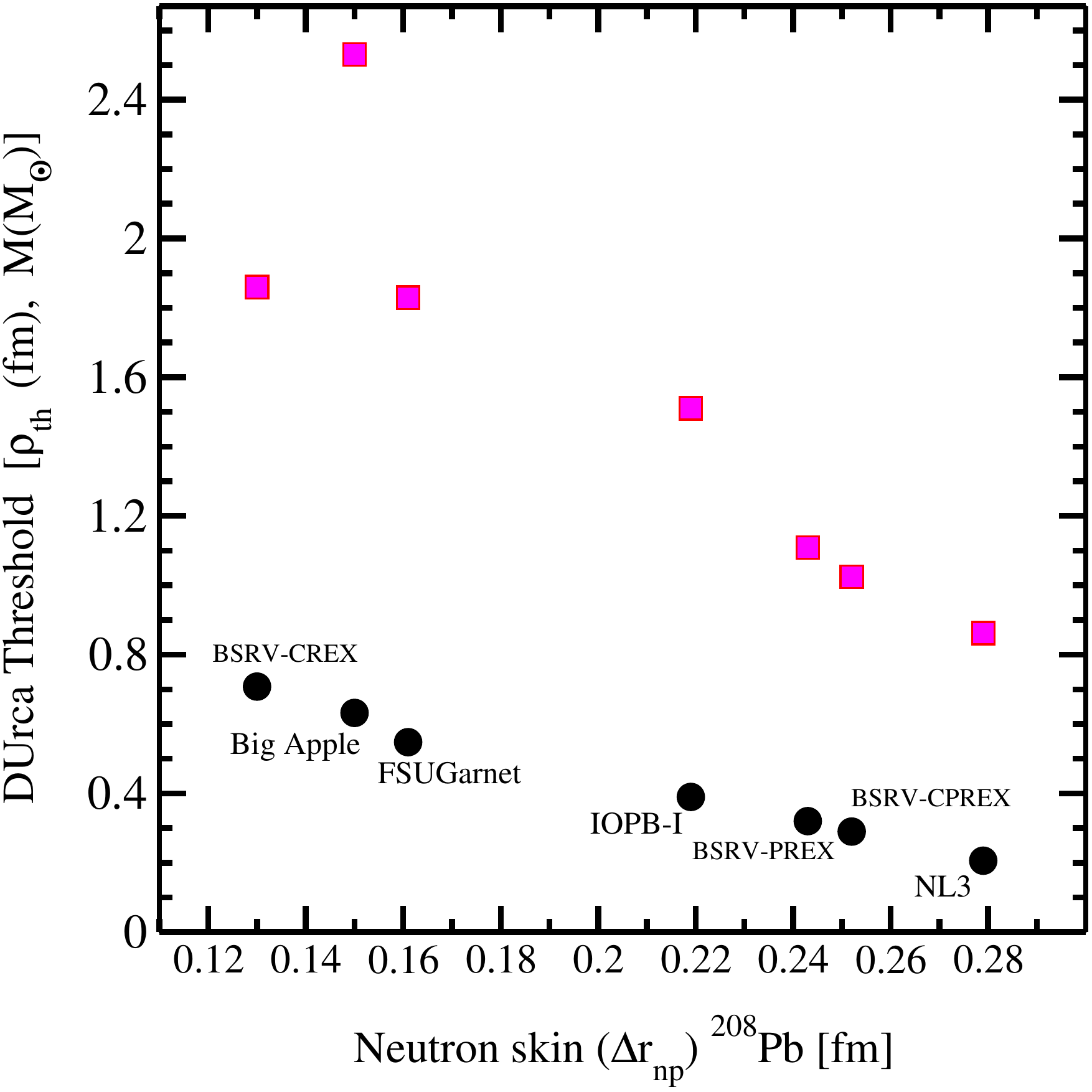}
\caption{\label{urca} (Color online) The thresholds for the onset of direct Urca Process in neutron stars. The threshold density is represented by the solid black circles and the corresponding neutron star mass for  such central density is depicted by solid magenta squares.  }
\end{figure}
%\clearpage
In Fig. \ref{urca} we plot the threshold central density (solid black circles) and corresponding neutron star mass (solid magenta squares)  as a function of neutron skin thickness for the onset of the direct Urca process for BSRV's parameter sets. Similar results for NL3, IOPB-1, FSUGarnet and Big Apple models are also displayed. It can be observed that the onset of the Urca process has a strong dependence  on neutron skin thickness of $^{208}$Pb and hence on symmetry energy coefficient. In particular, stiff symmetry energy as suggested by PREX-II favors large proton fractions that may trigger the onset of direct Urca process at lower central density \cite{Reed2021}.  It is evident from the figure that the onset of the direct Urca process is delayed with the decrease in the $\Delta r_{np}$ of the $^{208}Pb$. The  threshold density for the onset of the direct Urca process  for the BSRV-CREX model is   0.71 $fm^{-3}$. This high value is attributed to the small neutron skin thickness and  soft L for BSRV-CREX. The threshold density for the onset of the direct Urca process decreases from 0.71 $fm^{-3}$ to 0.21 $fm^{-3}$ as the value of $\Delta r_{np}$ for $^{208}$Pb increases from 0.13 fm (BSRV-CREX) to 0.28 fm (NL3).  The neutron star mass corresponding to threshold central density for the direct Urca process decreases from 2.53 to 0.86 ${M_{\odot}}$ for the parametrizations considered in the present work.\\
\subsection{Correlations amongst nuclear matter observables and model parameters}
\begin{figure*}
\centering
\includegraphics[trim=0 0 0 0,clip,scale=0.8]{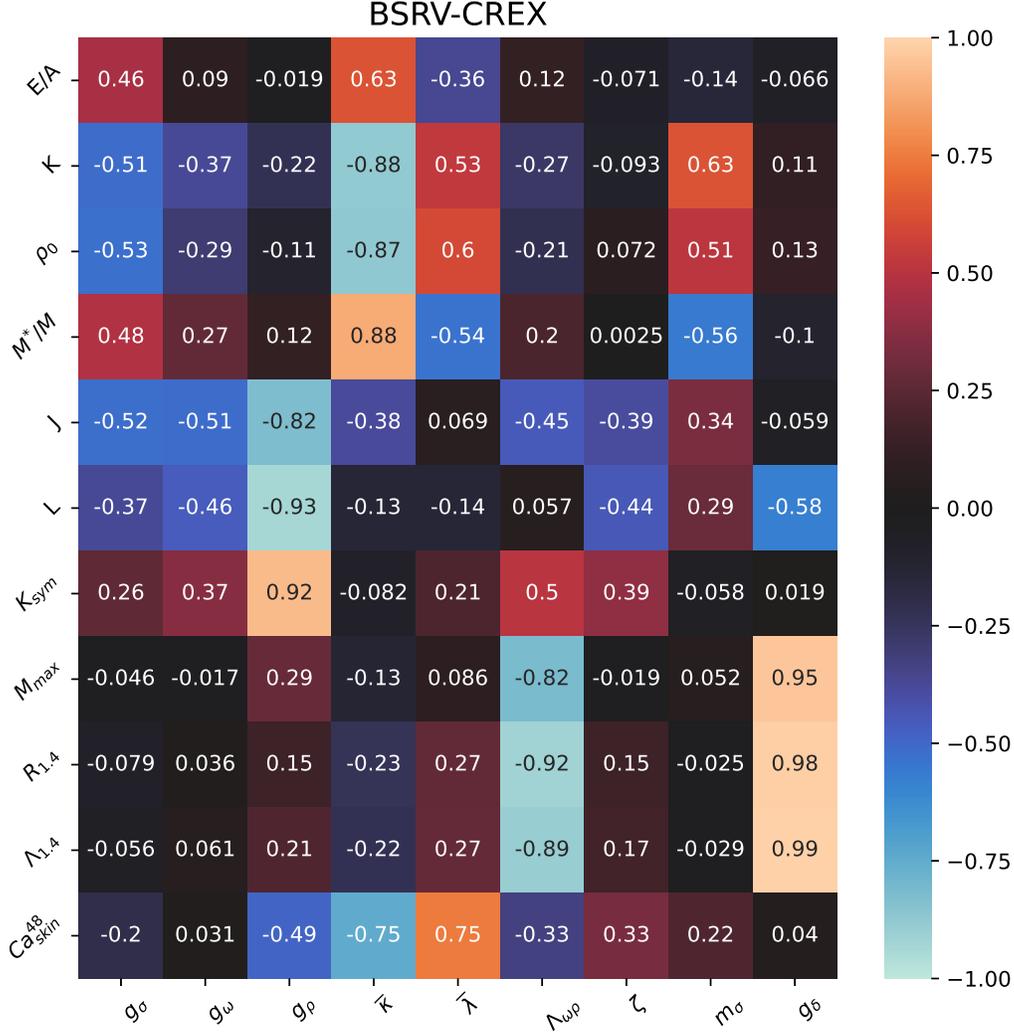}
\caption{\label{corr_prop_crex} (Color online) Correlation coefficients amongst neutron star observables  as well as the  bulk properties  of nuclear matter at the saturation density and model parameters for BSRV-CREX parametrization.}
\end{figure*}

\begin{figure*}
\centering
\includegraphics[trim=0 0 0 0,clip,scale=0.8]{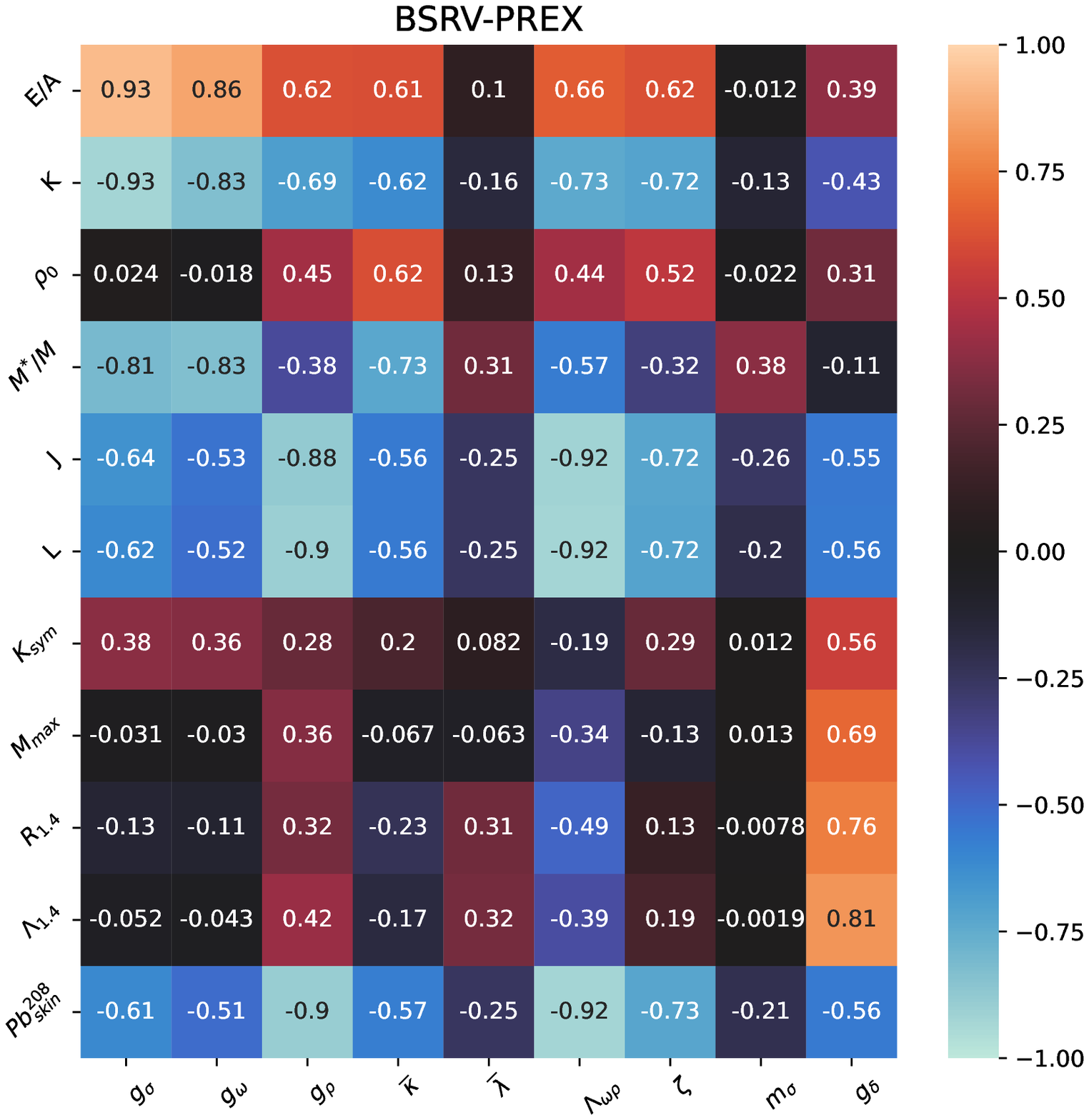}
\caption{\label{corr_prop_prex} (Color online) Same as Fig. \ref{corr_prop_crex}, but for BSRV-PREX paramerization.}
\end{figure*}

\begin{figure*}
\centering
\includegraphics[trim=0 0 0 0,clip,scale=0.8]{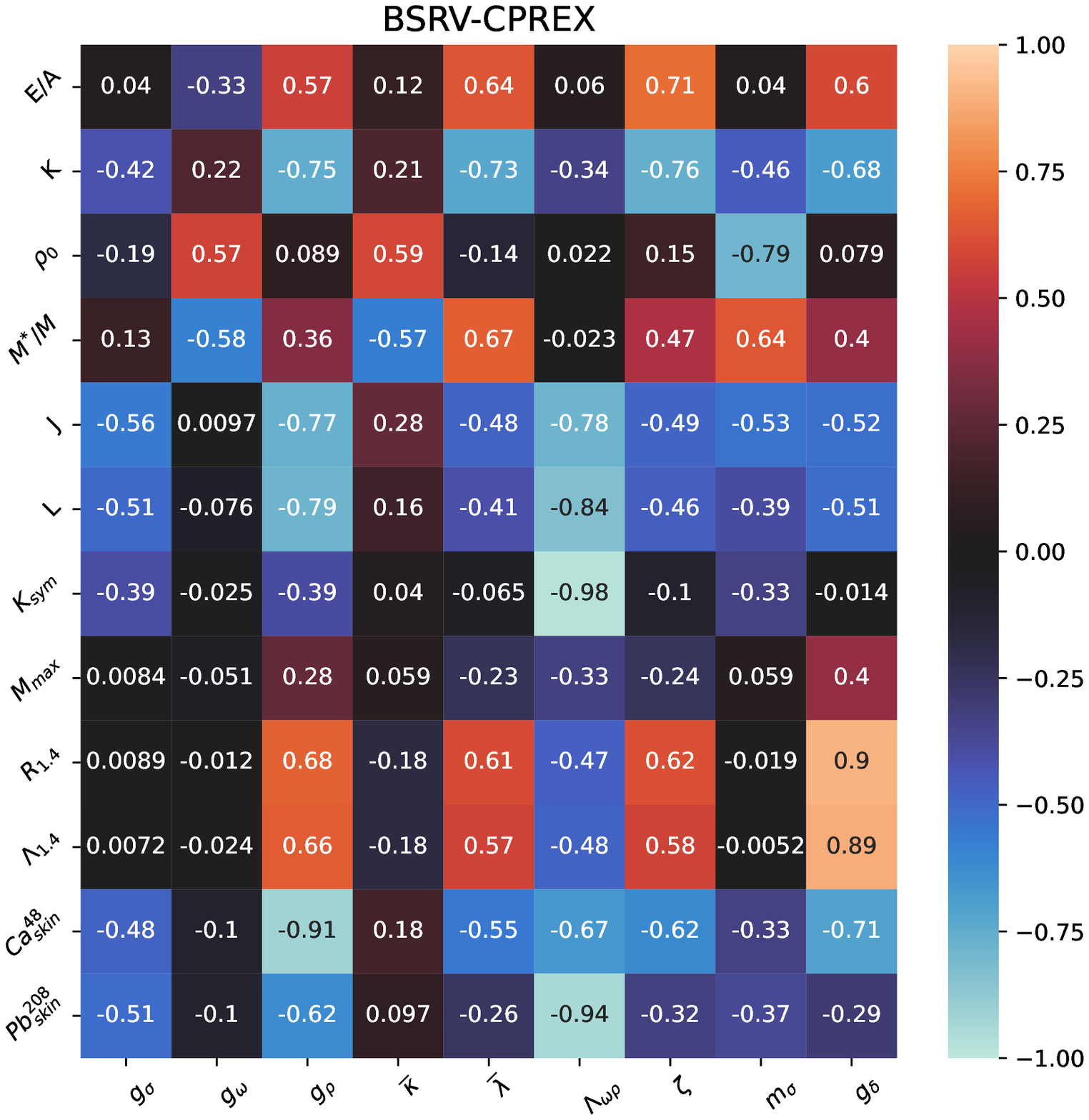}
\caption{\label{corr_prop_cprex} (Color online) Same as  Fig. \ref{corr_prop_crex}, but for BSRV-CPREX paramerization.}
\end{figure*}

In this sub-section, we discuss the  correlations  between nuclear matter observables and model parameters. In Fig. \ref{corr_prop_crex} to Fig. \ref{corr_prop_cprex}, we display the correlations of  bulk nuclear matter properties at saturation density and neutron star observables with the model parameters for BSRV's parametrizations.

For the BSRV-CREX model, the isoscalar bulk nuclear matter properties like  K, M*/M show strong correlations with isoscalar parameters $\overline{\kappa}$. It can also be observed from the Fig. \ref{corr_prop_crex} that the  symmetry energy coefficients (J), its density dependence (L) and curvature of symmetry energy $K_{sym}$  can be well constrained by the coupling parameter $g_{\rho}$ as suggested by their correlations. The neutron star observables like $M_{max}$, $R_{1.4}$ and $\Lambda_{1.4}$ display strong correlations with couplings $\Lambda_{\omega\rho}$ and $g_{\delta}$. A strong  correlation between $M_{max}$ and $\omega$-meson self-coupling parameter $\zeta$ is missing in the case of BSRV-CREX parameterization. The parameter $g_{\delta}$ might be  responsible for the large maximum mass of neutron star as it makes the EOS somewhat stiffer at high density and thus weakens the correlation between $M_{max}$ and $\zeta$.\\
For the BSRV-PREX model, the isoscalar bulk nuclear matter properties like  E/A, K and  M*/M show strong correlations with isoscalar parameters $g_{\sigma}$ and $g_{\omega}$. As expected, a strong correlation of  the isovector properties like  J,  L and $\Delta r_{np}$  for $^{208}$Pb  with isovector parameters $g_{\rho}$ and $\Lambda_{\omega\rho}$ is observed. The neutron star observables like $M_{max}$, $R_{1.4}$ and $\Lambda_{1.4}$ display good correlations with coupling  $g_{\delta}$ and weak correlations with $\zeta$. This suggests that the neutron star observables results from a competition  between $g_{\delta}$ and $\zeta$. For the BSRV-CPREX model, J, L and $\Delta r_{np}$  for $^{208}$Pb are very  well constrained by couplings $g_{\rho}$ and $\Lambda_{\omega\rho}$ as can be observed from their correlations. $K_{sym}$ shows strong dependence upon $\Lambda_{\omega\rho}$. The $\Delta r_{np}$  for $^{48}$Ca is very well constrained by $g_{\rho}$, $g_{\delta}$ and $\Lambda_{\omega\rho}$. The $R_{1.4}$ and $\Lambda_{1.4}$ is found to have strong correlation with $g_{\delta}$. These findings  are quite in harmony with the  results reported in Refs. \cite{Fattoyev2011,Chen2015}.
\section{Summary}\label{summary}
Three  relativistic interactions BSRV-CREX, BSRV-PREX and BSRV-CPREX  for the relativistic mean field model have been generated keeping in view the recently reported constraints on neutron skin thickness of $^{48}$Ca (CREX) and $^{208}$Pb nuclei by  PREX-II data. The precise measurements of neutron skin thickness give an opportunity to modify or readjust the coupling constants of RMF models without compromising the bulk properties of nuclear matter and neutron stars. The Lagrangian  density  for  the RMF model used in the present work  is  based on different non-linear, self and inter-couplings among isoscalar-scalar $\sigma$, isoscalar-vector $\omega_{\mu}$, isovector-scalar $\delta$ and isovector-vector $\rho_{\mu}$ meson fields and nucleonic Dirac field $\Psi$. The BSRV-CREX parametrization has been obtained by incorporating the recently measured neutron skin thickness $\Delta r_{np}$= $0.121\pm 0.026$ fm for $^{48}$Ca using the parity-violating electron scattering experiment \cite{Adhikari2022}. The parameters of the BSRV-PREX model have been searched by incorporating the recently measured neutron skin thickness $\Delta r_{np}$= $0.283\pm 0.071$ fm for $^{208}$Pb from the PREX-II  \cite{Adhikari2022} in our fit. The BSRV-CPREX parametrizations have been obtained by including  both the CREX and PREX-II data for neutron skin thicknesses for $^{48}$Ca and $^{208}$Pb nuclei in the fitting data.
The  BSRV's parameter sets  reproduce the ground state properties of the finite nuclei, bulk nuclear matter and also satisfy the constraints on mass and  radius along with dimensionless deformability ($\Lambda$) of a neutron star  from recent astrophysical observations \cite{Steiner2010,Annala2017,Abbott2018,Reed2021}.
 All the BSRV's parametrizations give an equally good fit to the finite nuclear properties. The Bulk nuclear matter properties obtained  are well consistent with the current empirical data \cite{Reed2021,Piekarewicz2014, Par2022}.
 The maximum gravitational mass and radius ($R_{1.4}$) of the neutron star lie in the range between 1.95$\pm$0.04 - 2.04$\pm$0.04 M$\odot$  and  12.66$\pm$0.39 - 13.41$\pm$0.71 km for BSRV's parameter sets respectively.  The value of $\Lambda_{1.4}$ which  lie in the range 525.84$\pm$151.48 - 682.57$\pm219.73$ for BSRV's parameterization also satisfies the GW170817 event \cite{Abbott2017,Abbott2018} and constraints obtained using Bayesian analysis and PREX-II reported in Refs. \cite{Chen2021,Abbott2017,Reed2021}. Covariance
analysis to measure the accuracy of the model parameters is also performed. This enabled us to estimate the statistical uncertainties
in the model parameters along with various correlations
amongst the nuclear matter observables and parameters.\\
 CREX and PREX-II data on neutron skin thickness has opened an important new perspective to constrain the density dependence of symmetry energy and  energy density functionals and having a considerable effect on the isovector channel. The BSRV-CREX parameterization obtained by incorporating CREX data provides the smaller value of symmetry energy, its slope at saturation density and neutron skin thickness  are significantly very small. This may be due to the large value of coupling $g_{\delta}$  which seems to play an important role during the calibration procedure of model parameters and is responsible for soft and stiff behaviour  of symmetry energy in medium and high density regime.  But the radius and  tidal deformability of 1.4$M_{\odot}$ neutron star reveals some tension with the revised limit of $\Lambda_{1.4}\le 580$ \cite{Abbott2018}. The BSRV-PREX parametrization obtained keeping in view the PREX-II data suggests stiff symmetry energy and  stiff EoS. For BSRV-PREX and BSRV-CPREX models, the tidal deformability of 1.4$M_{\odot}$ neutron star has some overlap  with the revised limit of $\Lambda_{1.4}\leq $ 580 within  1$\sigma$ uncertainty reported in Ref.\cite{Abbott2018}. BSRV-CREX gives soft density dependence of symmetry energy whereas BSRV-PREX and BSRV-CPREX provide its stiff value. The analysis of the present work shows that no consistent conclusion from the theoretical side could be obtained when using recently measured CREX and PREX-II data. We are hoping that novel experimental studies are necessary to resolve this conflict or discrepancies.
\begin{acknowledgments}
Author(s) are thankful to Himachal Pradesh University for providing the computational facility. BKA acknowledges partial support from the SERB, Department of science and technology, Govt. of India with grant numbers SIR/2022/000566 and CRG/2021/000101 respectively. SK is highly thankful to CSIR-UGC (Govt. of India) for providing financial assistance (NTA/ 211610029883 dated
19/04/2022) under Junior/Senior Research Fellowship scheme.

 \end{acknowledgments}
\bibliography{Ref}
 \bibliographystyle{unsrt}
 \end{document}